\documentclass[runningheads]{llncs}
\usepackage[T1]{fontenc}
\usepackage{graphicx}
\usepackage{xcolor}
\usepackage{subcaption}
\usepackage[colorlinks,linkcolor=blue,citecolor=blue,urlcolor=blue]{hyperref}

\usepackage{marvosym}
\usepackage{amsmath,amsfonts,amssymb}
\usepackage{mathrsfs}
\usepackage{extarrows}
\usepackage{textcomp}
\usepackage{stmaryrd}
\usepackage{pifont}

\usepackage[ruled,vlined,linesnumbered]{algorithm2e}

\usepackage{booktabs}
\usepackage{makecell}
\usepackage{threeparttable}

\usepackage[shortlabels]{enumitem}
\setlist{nolistsep} 

\newenvironment{defi}{\begin{definition}}{\end{definition}}

\newenvironment{lem}{\begin{lemma}}{\end{lemma}}

\begin{document}
\title{Skyline Community Search over Edge-Attributed Bipartite Graphs}
\titlerunning{Skyline Community Search over Edge-Attributed Bipartite Graphs}

\author{
Fangda Guo\inst{1}\textsuperscript{*}\textsuperscript{(\Letter)}\orcidID{0000-0003-2401-6499}\and
Xuanpu Luo\inst{1}\textsuperscript{*}\orcidID{0009-0008-4507-1466} \and
Shiyuan Xu\inst{2}\textsuperscript{(\Letter)} \orcidID{0000-0001-9076-1695} \and
Haowen Gao\inst{1,3}\textsuperscript{(\Letter)} \orcidID{0009-0007-9877-6163}\and
Yanghao Liu\inst{1,3}\orcidID{0009-0001-7979-2885} \and
Huawei Shen\inst{1}\orcidID{0000-0002-1081-8119} \and
Xueqi Cheng\inst{1}\orcidID{0000-0002-5201-8195}
}

\authorrunning{F. Guo et al.}

\institute{
State Key Lab of AI Safety, Institute of Computing Technology, CAS, Beijing, China \\
\email{\{guofangda, seuviplxp, shenhuawei, cxq\}@ict.ac.cn} \and
School of Computing and Data Science, The University of Hong Kong, Hong Kong, China \\
\email{syxu2@cs.hku.hk} \and
University of Chinese Academy of Sciences, Beijing, China \\
\email{\{gaohaowen23s, liuyanghao19s\}@ict.ac.cn}
}

\maketitle
\begingroup
\renewcommand{\thefootnote}{\fnsymbol{footnote}}
\footnotetext[1]{Equal contribution.}
\endgroup
\begin{abstract}
Bipartite graphs, modeling relationships between two types of entities, are widely used in practical applications. Community search, a fundamental problem in bipartite graphs, has gained significant attention. However, existing studies focus on measuring structural cohesiveness between vertex sets while either ignoring edge attributes or considering only one-dimensional importance. In this paper, we introduce a novel community model, named edge-attributed skyline community (ESC), which preserves structural cohesiveness and captures the inherent dominance of multi-dimensional edge attributes in bipartite graphs. To search for ESCs, we developed an efficient peeling algorithm that iteratively deletes edges with the minimum attribute in each dimension. Additionally, we devised an expanding algorithm to reduce the search space and speed up the filtering of unpromising vertices using a proven upper bound. Extensive experiments on large-scale real-world datasets demonstrate the efficiency, effectiveness, and scalability of our approach. A case study compared with prior arts demonstrates that our design improves the precision and diversity of results.

\end{abstract}
\section{Introduction}
Numerous real-world networks consist of community structures, making their discovery crucial for network analysis \cite{bai2024secmdp,xu2023effective,yuan2025indexbase,11184653}. Community search aims to find the densest connected subgraphs containing specific input queries. It has garnered significant attention from database professionals in recent years \cite{xu2024model,xu2025lattice,meng2024survey}.
Traditional search focuses primarily on the network topology \cite{xu2025multi,11195061,chen2023index,xu2025lattice2}. However, in practical applications, it needs to consider not only the connections but also their attribute characteristics, as these can enhance the cohesion of identified communities \cite{guo2023fededb,guo2023verifying}. This makes community search over attributed graphs particularly significant.
Most current research addresses community constraints in terms of structure and attributes but typically only for scenarios involving a single type of entity \cite{luo2020efficient,guo2021multi}. However, in scenarios involving two different types of entities, such as collaboration networks \cite{ley2002dblp}, customer-product networks \cite{wang2006unifying,xu2024post}, and user-location networks \cite{chen2021efficiently}, the focus tends to be on structural cohesiveness, often neglecting attributes. Approaches like the $(\alpha, \beta)$-core \cite{ding2017efficient}, bitruss \cite{sariyuce2018peeling,wang2020efficient}, and biclique \cite{zhang2014finding,wang2022efficient} exemplify this trend by primarily concentrating on structure.

Some researchers considered one-dimensional importance for attribute information \cite{wang2023Discovering}, but interactions often involve multiple dimensions. They missed the multidimensional aspects and fail to capture all cohesive communities. We aim to develop a model using both structural and multi-dimensional attribute information to identify all cohesive communities containing the queries. By designating $U_2$ as the query vertex and setting structural constraints of $\alpha=\beta=2$, we extract two mutually exclusive communities\footnote{Strictly speaking, communities may overlap.} with the largest connected subgraphs based on the multidimensional attributes.

In this study, our goal is to identify vertices in the bipartite graph that are structurally closest to the query vertex and have similar attribute values, along with the corresponding edges, to find high-importance communities that meet these criteria.
To achieve this, we propose a skyline community search problem for bipartite graphs with multi-dimensional edge attributes, named Edge-attributed Skyline Community Search Problem \textbf{(ESCS-Problem)}, along with its model termed Edge-attributed Skyline Community \textbf{(ESC)}.

Our initial effort targets bipartite graphs with multidimensional edge attributes, presenting a new challenge. A recent study \cite{li2018skyline} focuses on communities with multidimensional attributes in simple graphs, while another \cite{wang2021efficient} deals with single-dimensional attributes, making it a subset of our scope. Additionally, the design \cite{2021Pareto} using Pareto optimality for one-dimensional weights differs in context and approach. Other works \cite{zhou2023influential,liu2023significant} on attribute community discovery in heterogeneous graphs handle only one-dimensional attributes and vertex-associated values, unsuitable for our scenario. We face challenges in balancing structural compactness with multidimensional attribute advantages, marking a creative endeavor.

\noindent \textbf{Extensive Applications for ESC:}

\textbf{(1) Personalized Recommendation Systems.} Our ESC model surpasses current solutions by integrating structural cohesion and multi-dimensional edge attributes. Unlike existing models that are constrained to either structure or single attributes, ESC captures a diverse range of user behaviors. For movie recommendations, it considers ratings, likes, and viewing duration. On social platforms, it refines blogger recommendations by simultaneously analyzing likes, follows, and comments.

\textbf{(2) Team Formation.} The ESC model enhances team assembly by balancing structural cohesion and multi-dimensional contributions. Unlike models that focus solely on experience, ESC evaluates factors such as impact, involvement duration, and deliverable volume to form teams with both deep expertise and relevant project experience. It ensures that the teams meet multiple objectives without making trade-offs, which is often a limitation of existing methods that may overlook the importance of maintaining such a balance.

\textbf{(3) Anomaly Detection.} The ESC model enhances fraud detection by utilizing multi-dimensional bipartite graphs, surpassing models that rely on single anomaly indicators. By jointly assessing factors such as financial scale, affected users, and severity, it ensures robust risk identification. Its ability to detect diverse yet structurally significant anomalies strengthens fraud prevention, addressing gaps left by traditional models.

Our contributions are summarized as follows:

\begin{itemize}
\item \textbf{Novel community model.} We introduce a new model for community search in bipartite graphs, considering both structural cohesion and multi-dimensional edge attributes.
\item \textbf{New algorithms.} We propose an efficient peeling algorithm that removes edges with the minimum attribute and an expanding algorithm to reduce search space and speed up filtering. We also give lemmas to improve search performance, and verify computational complexity and completeness.
\item \textbf{Extensive experiments.} We conduct extensive experiments over six real-world datasets to demonstrate the high efficiency and scalability of our proposed algorithms. We also give an extensive case study, demonstrating our ESC model's capability to match dense subgraphs of bipartite graphs with greater accuracy and personalization.
\end{itemize}

\section{Problem Statement}

\begin{defi}
\emph{\textbf{(Bipartite Graph.) }}
A bipartite graph $G = (U,L,E,$ $X)$ is a graph whose vertices can be divided into two disjoint sets such that every edge connects a vertex from one set to a vertex from the other set. Note that $U$ and $L$ denote the sets of upper and lower vertices, respectively, and $E$ represents the set of edges in $G$. The edges in $G$ have $d$ numerical attributes, which are denoted by $X_i$, where $X_i \in X$ and $X_i$ is a $d$-dimensional vector. Let $|*|$ denote the modulo operation. We have $|U|+|L|=n$ and $|E|=m$.
\end{defi}

\begin{defi}
\emph{\textbf{(($\alpha$, $\beta$)-core.) }}
Given a bipartite graph $G=(U, L, E, $ $X)$, an ($\alpha$, $\beta$)-core is a connected subgraph where the degree of vertices in the upper layer is at least $\alpha$, and the degree of vertices in the lower layer is at least $\beta$.
\end{defi}

\begin{defi}\label{def1}
\emph{\textbf{(Significance.) }}
Given a bipartite graph $G=(U,L,E,X)$ with edge attributes, we define the significance of $G$ on the $i^{th}$ dimension (for $i = 1, 2,\cdots, d$) as: $f_i(G)=\mathop{\min}_{e\in|E|}X_i^e$.
\end{defi}

The significance bounds an attribute dimension, with higher values indicating stronger cohesion. In a user-movie network, edge attributes represent ratings, similar ratings suggesting community membership. The final output subgraph, ESC, must not be dominated-any lower significance implies non-optimality.

\begin{defi}
\emph{\textbf{(Domination.) }}
Given two bipartite graphs $G = (U,L,E,X)$ and $G' = (U',L',E',X')$ , if $f_i(G)\leq f_i(G')$ (for $i = 1, 2, \cdots , d$) and there exists $f_i(G) < f_i(G')$ for certain $i$, we call $G'$ dominates $G$, denoted by $G \preceq G'$.
\end{defi}

\begin{defi}
\label{def:ESC}
\emph{\textbf{(ESC.) }}
Given a bipartite graph $G = (U,L,E,X)$, $\alpha$, $\beta$ and query vertex $q$, where $X$ is the multi-dimensional significance on edges \cite{li2018skyline,wang2021efficient,2021Pareto}. An ESC is a connected subgraph $H = (U_H,L_H,E_H,X_H)$ of $G$, where:
\begin{enumerate}
\item[1)]
\emph{Cohesive property:} $H$ is an ($\alpha$ , $\beta$)-core contains $q$. It ensures that the connected subgraph contains the query and follows the ($\alpha,\beta$)-core structure.
\item[2)]
\emph{Skyline property:} it does not exist a subgraph $H'$ of $G$ such that $H'$ is an ($\alpha$, $\beta$)-core and $H \prec H'$. It guarantees that the subgraph is not dominated by others, making it an optimal solution.
\item[3)]
\emph{Maximal property:} it does not exist a subgraph $H'$ of $G$ such that (1) $ H'$ is an ($\alpha$, $\beta$)-core, (2) $H'$ contains $H$, and (3) $f_i(H) = f_i(H')$ for all $i \in [1, d]$. It ensures the largest possible subgraph under constraints, maximizing member inclusion.
\end{enumerate}
\end{defi}

\textbf{ESCS-Problem:} Given a bipartite graph $G$, a query vertex $q$, and parameters $\alpha$ and $\beta$ for structural constraints, the problem is aimed to search all ESCs containing $q$, where each ESC $H$ cannot be dominated by any other ESC $H'$.

\begin{figure}[ht]
\centering
\includegraphics[width=0.9\columnwidth]{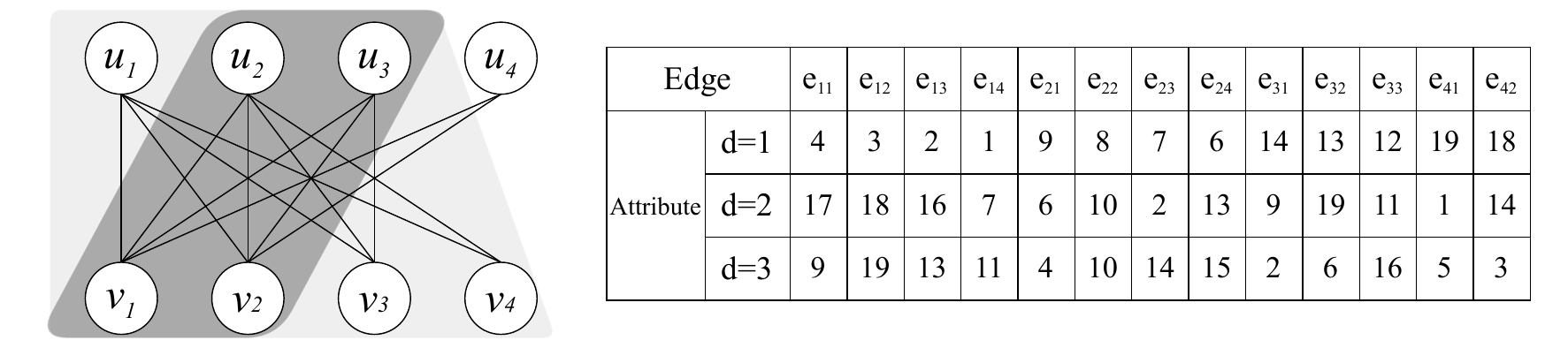}
\caption{Example of localized IMDB bipartite graph.}
\label{fig1}
\end{figure}

\begin{example}
To illustrate the definition of ESC, Fig.~\ref{fig1} depicts a bipartite graph segment from the IMDB network with three-dimensional edge attributes. Here, "$e_{ij}$" signifies the edge that connects node $u_i$ with node $v_j$. We concentrate on two-dimensional attributes for simplicity. Let $U_2$ be the query vertex and $\alpha=\beta=2$, leading to the identification of two ESCs, $H_1$ and $H_2$, each highlighted within a dotted circle.
$H_1$ is composed of the vertices $U_2,U_3,$ $V_1,V_2$, and $H_2$ includes ${U_1,U_2,U_3,V_1,V_2,V_3,V_4}$. Both $H_1$ and $H_2$ incorporate $U_2$ and are defined as ($2$, $2$)-cores. The significance of $H_1$ is captured by $f(H_1)=(8,6)$, and for $H_2$, it is $f(H_2)=(1,7)$. This configuration ensures that they are impervious to dominance by other ($2$, $2$)-cores based on their significance.
\end{example}

\section{Peeling-based Algorithm}

We define the ESC and propose algorithms to detect it in large-scale bipartite graphs by combining ($\alpha$, $\beta$)-core constraints with significance measures. The size gap between the ESC and the original graph impacts efficiency, so we introduce two approaches: a peeling algorithm for large disparities and an expanding algorithm for smaller ones. The peeling method iteratively removes edges that violate conditions, making it efficient when the ESC is much smaller. Prioritizing edge removal based on multidimensional significance, our approach adapts to different dimensions, recognizing their hierarchical structure for optimal processing.

\subsection{Peeling for $d = 1$}

In simple bipartite graphs with one-dimensional attributes, the algorithm follows a straightforward approach. It first sorts edges in ascending order of significance, then iteratively removes the least significant edge until the structure no longer maintains an ($\alpha$, $\beta$)-core or the queried vertex is excluded. The resulting ($\alpha$, $\beta$)-core
is called an ESC. In the operational phase, peeling one edge at a time is inefficient. To address this, we've introduced \textsf{DFSCom}, as depicted in Alg. \ref{alg:algorithm1}, which recursively traverses adjacent vertices to ensure all structural constraints are met. Once these criteria are satisfied, the algorithm ends, removing all non-compliant vertices and their edges.

\begin{algorithm}[ht]
\caption{PDim1}
\label{alg:algorithm1}
\KwIn{$G, \alpha, \beta, q, I, d, F$}
\KwOut{$f_d$}
\BlankLine
$G \gets G \setminus \{e | e \text{ violates } I\}$;
$G(q) \gets \text{maximal } (\alpha,\beta)\text{-core containing } q$;
$f_d \gets f_d(G(q))$;

\While{$G(q) \neq \emptyset$}{
$w_{min} \gets \min_{d\text{-dim}} G(q)$;
\ForEach{$(u,v) \in G(q)$}{$visited(u,v) \gets 0$;}

\ForEach{$(u,v) \in G(q) \text{ with } w(u,v) = w_{min}$}{
\lIf{$(u,v) \in F$}{\textbf{break}}
$S \gets \emptyset$, $visited(u,v) \gets 1$;

\lIf{$\text{DFSCom}(u,q,\alpha,\beta,S,F) = 0$}{\textbf{break}}
\lIf{$\text{DFSCom}(v,q,\beta,\alpha,S,F) = 0$}{\textbf{break}}

$G(q) \gets G(q) \setminus \{(u,v)\}$;
$f_d \gets \max(f_d, f_d(G(q)))$;
}
}
\Return{$f_d$};

\SetKwFunction{FMain}{DFSCom}
\SetKwProg{Fn}{Function}{:}{end}
\Fn{\FMain{$u, q, \alpha, \beta, S, F$}}{
\lIf{$u=q \land deg(u,G(q))=\alpha$}{\Return{0}}
Update $deg(u,G(q))$;

\uIf{$deg(u,G(q)) < \alpha$}{
\ForEach{$v' \in N(u) \text{ with } visited(u,v')=0$}{
$visited(u,v') \gets 1$;
$S \gets S \cup \{(u,v')\}$;
\lIf{$(u,v') \in F \lor \text{DFSCom}(v',q,\beta,\alpha,S,F)=0$}{
$G(q) \gets G(q) \cup S$;
\Return{0}
}
}
$S \gets \emptyset$;
}
\Return{1};
}
\end{algorithm}

\noindent \textbf{Completeness:} The integrity of the algorithm is elucidated as follows. Assuming there exists another ESC $H'$ that contains edges not present in the current ESC $H$. It must satisfy at least one of the following two conditions: (1) it contains edges with significance smaller than $f(H)$, which contradicts the definition of $f(H')$ and is not possible; or (2) it contains edges with significance greater than $f(H)$, which were removed because the two vertices they connect do not satisfy the structural constraints. When $H'$ adds these edges to the community, it must also add edges with significance smaller than $f(H)$ in order to satisfy the ($\alpha$, $\beta$)-core constraint, which contradicts the definition of $f(H')$ and is not possible. Therefore, it is not possible to have another ESC.

\noindent \textbf{Time Complexity:} Assuming that the number of edges in $G$ is $m$ and of vertices is $n$. Generally speaking, the time complexity of obtaining ($\alpha$, $\beta$)-core from the graph is $O(m+n)$. The basic idea involves removing one edge in each iteration and checking for the presence of ($\alpha$, $\beta$)-core. Consequently, the time complexity is $O(m \cdot(n+m))$. In contrast, the approach with pruning strategies, on average, results in each vertex having $m/n$ neighbors. Therefore, the average algorithm can reduce $m/n$ checks, leading to a time complexity of $O((m-m/n) \cdot(n+m))$.

\subsection{Peeling for $d = 2$}

As edge significance dimensionality grows, a backtracking-based dimensionality reduction strategy decomposes two-dimensional significance into one-dimensional sets, enabling existing methods to tackle new problems. Based on Lemma \ref{lemma1}, we propose the Alg. \ref{alg:algorithm2}.

\begin{lem}\label{lemma1}
Without loss of generality, assuming that there exist $n$ ESCs $H_i$, $(i = 1,2,...,n)$ in the bipartite graph $G$ with two dimensions. Then, the significance of each dimension corresponding to different ESCs can be related as:
\begin{equation}
\begin{aligned}
f_1(H_1) &< f_1(H_2) &< ... &< f_1(H_i) &< ... &< f_1(H_n)\\
f_2(H_1) &> f_2(H_2) &> ... &> f_2(H_i) &> ... &> f_2(H_n)
\end{aligned}
\end{equation}
\end{lem}

\begin{algorithm}[ht]
\caption{PDim2}
\label{alg:algorithm2}
\KwIn{$G, \alpha, \beta, q, I, F$}
\KwOut{$R$}
\BlankLine
$f_2\gets$ PDim1($G, \alpha, \beta, q, I, 2, F$);

$R\gets \varnothing$;

\While{\textnormal{$f_2 > 0$}}{
$I'\gets I.update( \{X_2^i \geq f_2\})$;

$f_1\gets$ PDim1($G, \alpha, \beta, q, I', 1, F$);

$R \gets R \cup \{(f_1, f_2)\}$;

$I'\gets I.update( \{X_1^i > f_1\})$;

$f_2\gets$ PDim1($G, \alpha, \beta, q, I', 2, F$);
}
\Return{$R$};
\end{algorithm}

\begin{figure}[ht]
\centering
\begin{subfigure}{0.3\columnwidth}
\includegraphics[width=\linewidth]{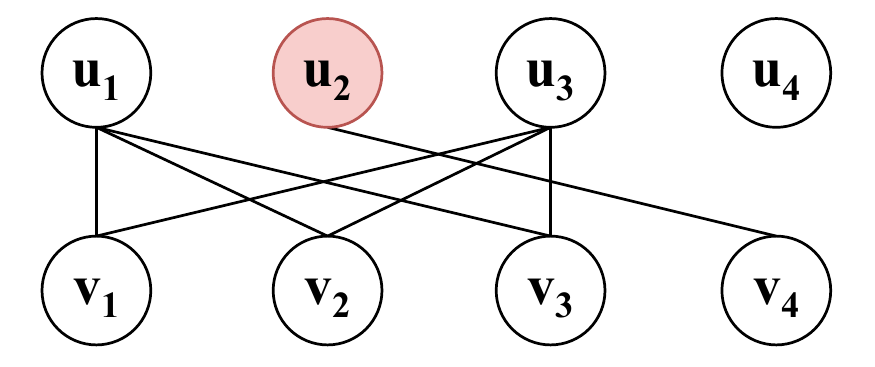}
\caption{}
\label{figpeel2fig:case_a}
\end{subfigure}
\hfill
\begin{subfigure}{0.3\columnwidth}
\includegraphics[width=\linewidth]{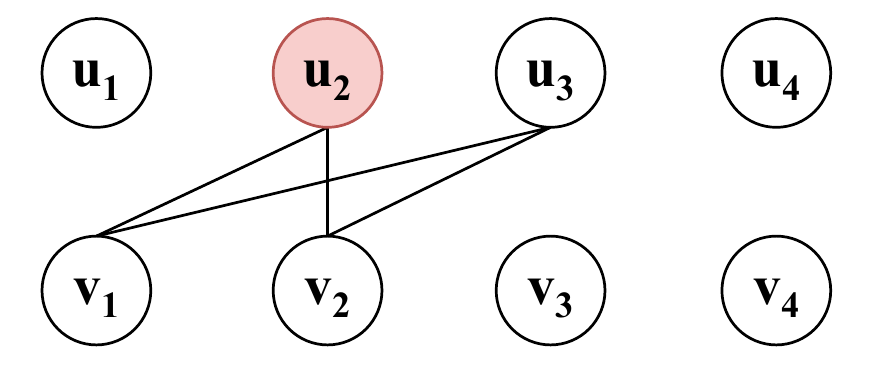}
\caption{}
\label{figpeel2fig:case_b}
\end{subfigure}
\hfill
\begin{subfigure}{0.3\columnwidth}
\includegraphics[width=\linewidth]{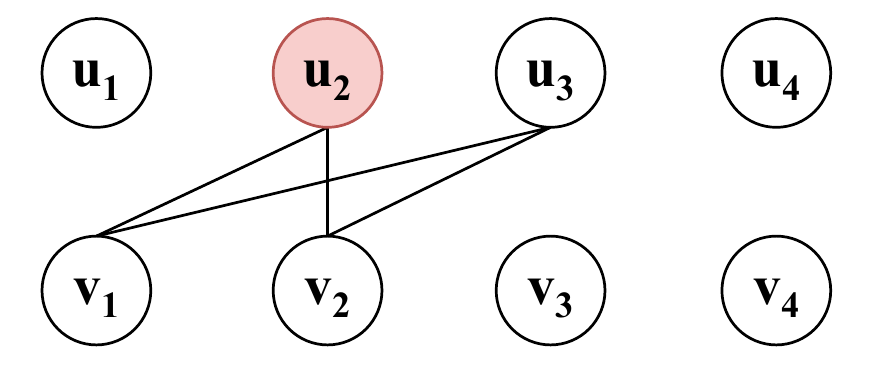}
\caption{}
\label{figpeel2fig:case_c}
\end{subfigure}
\caption{An example of the peeling algorithm for $d=2$.}
\label{figpeel2}
\end{figure}

\begin{example}
Referring to Fig.~\ref{fig1}, it is invoked on the significance of the second dimension to obtain an upper bound of $f_2(H_1)$. Then, all edges in the original graph $G$ with $X^e_2 < f_2(H_1)$ are removed. Algorithm expending ($d$=1) is applied to the attribute values of the first dimension, yielding $f_1(H_1)=1$, resulting in the first $H_1=(1,7)$ as depicted in Fig~\ref{figpeel2}(a). All edges in the original graph $G$ with $X^e_1 > f_1(H_1)$ are retained, as shown in Fig~\ref{figpeel2}(b). Algorithm expending($d$=1) is called upon again to update $f_2(H_2)$. This iterative process continues until $f_2=0$. At this point, as shown in Fig~\ref{figpeel2}(c), another ESC $H_2=(8,6)$ is obtained.
\end{example}

\noindent \textbf{Completeness:}
Suppose an undiscovered ESC $H$. According to Lemma~\ref{lemma3}, we have $f_1(H_i) < f_1(H) < f_1(H_{i+1})$ and $f_2(H_j) > f_2(H) >f_2(H_{j+1})$. Assuming $i<j$, we have $f_1(H_i) < f_1(H) < f_1(H_{j})$. Also, since $f_2(H_j) > f_2(H)$, $H_j$ dominates $H$, thus $H$ cannot exist. The same can be proven when $i>j$. Therefore, Algorithm expanding($d=2$) is complete.

\noindent \textbf{Time Complexity:} There will be $\epsilon$ rounds of iterations if there are $\epsilon$ ESCs. The one-dimensional peeling algorithm must be called twice in each iteration, resulting in a time complexity of $O(\epsilon \cdot(m-m/n)\cdot(n+m))$.

\subsection{Peeling for $d = 3$}

We implement an alternate Algorithm expending ($d=3$).
For obtaining $f_3(H_i)$, the process of being transferred function \textsf{CandSig} is as follows: in the process of gradually removing the edge with the minimum $X_d$, recording each $X_d$ that makes the subgraph contain ($\alpha$,$\beta$)-core. These recorded values constitute all possible significance for ESCs on $d_{th}$ dimension.
Then, we iterate through the set $V$ of all possible values $v_i\in V$ during each iteration, edges with third-dimensional attribute values$X_3^e\leq v_i$ are removed.

\begin{algorithm}[ht]
\caption{CandSig}
\label{alg:algorithm3}
\KwIn{$G, \alpha, \beta, q, d$}
\KwOut{$T_d$}
\BlankLine
$T_d \gets \emptyset$;
$G(q) \gets \text{maximal } (\alpha,\beta)\text{-core containing } q$;

\While{$G(q) \neq \emptyset$}{
$f_d \gets \min_{d\text{-dim}} G(q)$;
$T_d \gets T_d \cup \{f_d\}$;

Remove $e = (u,v) \in G(q)$ with $X_d^e = f_d$;

\text{DFS}($G(q)$, $u$, $\alpha$, $\beta$);
\text{DFS}($G(q)$, $v$, $\beta$, $\alpha$);

$G(q) \gets \text{maximal } (\alpha,\beta)\text{-core containing } q$;
}
\Return{$T_d$};

\SetKwFunction{FMain}{DFS}
\SetKwProg{Fn}{Function}{:}{end}
\Fn{\FMain{$G, u, \alpha, \beta$}}{
\If{$deg(G,u) > \alpha$}{
\ForEach{$v \in N(u)$}{
$G \gets G \setminus \{(u,v)\}$;
\text{DFS}($G, v, \beta, \alpha$);
}
}
}
\end{algorithm}

\begin{algorithm}[ht]
\caption{PDim3}
\label{alg:algorithm4}
\KwIn{$G, \alpha, \beta, q, I, F$}
\KwOut{$R$}
\BlankLine
$F_3 \gets \text{CandSig}(G, \alpha, \beta, q, 3)$;
$R, S \gets \emptyset$;
Sort $F_3$ descending;

\ForEach{$f_3 \in F_3$}{
$e \gets \{e \in G \mid X_3^e = f_3\}$;
$F \gets F \cup e$;

\lIf{$(X_1^e, X_2^e) \preceq S$}{\textbf{continue}}

$I' \gets I \cup \{X_3 \geq f_3\}$;
$T \gets \text{PDim2}(G, \alpha, \beta, q, I', F)$;

$S \gets S \cup T$;
\ForEach{$(f_1,f_2) \in T$}{
$R \gets R \cup \{(f_1,f_2,f_3)\}$;
}
}
\Return{$R$};
\end{algorithm}

\noindent \textbf{Completeness:}
If there is any other $f_3(H_i)$ present, it is impossible to form an ($\alpha$, $\beta$)-core. When fixing the $f_3(H_i)$, assume the existence of an ESC $H'$ in the subgraph on the other two dimensions that Algorithm expending($d$=2) did not find. As $H'$ must be dominated by the already identified ESCs, contradicting the definition of ESC. Therefore, the algorithm does not overlook ESC solutions. 

\noindent \textbf{Time Complexity:} To obtain all possible $f_3$ of ESCs, it requires the time complexity of $O((m-m/n)\cdot (n+m))$. Algorithm~\ref{alg:algorithm2} needs to be invoked in each iteration, and the upper bound on the number of iterations is $m$. Consequently, the final time complexity is $O(m\cdot(m-m/n)\cdot(n + m))$.

\subsection{Universal Peeling}

For high-dimensional cases ($d>3$), we propose an expanding ($d>3$) algorithm based on dimensionality reduction and backtracking. It collects all possible values of $f_d(H_i)$ as $F_d$ and iterates over each $f_d$ in $F_d$. In each iteration, edges with $d_{th}$ dimensional significance $X_d^e\leq f_d$ are removed, and the $(d-1)_{th}$ peeling algorithm is applied, reducing dimensionality until $d=3$. Then, it retrieves all relevant ESCs. A pruning strategy improves efficiency by preventing redundant calculations. For each ESC, $f_d$ ensures that edge $e$ with $X_d^e=f_d$ is present, setting upper bounds on other dimensions based on $e$’s significance. If these bounds are dominated by previously found ESCs, no ESC corresponds to $f_d$, accelerating computation.

\begin{algorithm}[ht]
\caption{UniPDim}
\label{alg:algorithm5}
\KwIn{$G, \alpha, \beta, q, I, F, d$}
\KwOut{$R$}
\BlankLine
\lIf{$d=3$}{\Return{\text{PDim3}(G, $\alpha$, $\beta$, q, I, F)}}
$F_d \gets \text{CandSig}(G, \alpha, \beta, q, d)$;
$R \gets \emptyset$;
Sort $F_d$ descending; $S \gets \emptyset$;

\ForEach{$f_d \in F_d$}{
$e \gets \{e \in G \mid X_d^e = f_d\}$;
$F \gets F \cup e$;

\lIf{$(X_1^e, \dots, X_{d-1}^e) \preceq S$}{\textbf{continue}}

$I' \gets I \cup \{X_d \geq f_d\}$;
$T \gets \text{UniPDim}(G, \alpha, \beta, q, I', F, d-1)$;
$S \gets S \cup T$;

\ForEach{$(f_1, \dots, f_d) \in T$}{
$R \gets R \cup \{(f_1, \dots, f_d)\}$;
}
}
\Return{$R$};
\end{algorithm}

\noindent \textbf{Time Complexity:} The time complexity of Algorithm~\ref{alg:algorithm5} can be derived from Algorithm~\ref{alg:algorithm4} since the entire algorithm operates through iterative recursion. Let the upper bound on the number of iterations in each round be $m$, for a bipartite graph with $d$-dimensional attributes, it requires at most $m^{d-1}$ iterations. Therefore, the time complexity is $O(m^{d-1}\cdot(m-m/n)\cdot(n + m))$.

\section{Expanding-based Algorithm}

The peeling-based algorithm is proficient and effective in handling large-scale graphs, albeit with intrinsic limitations. Intuitively, each search requires loading the entire graph to execute the peeling process, which becomes excessively resource-intensive when dealing with smaller graph scales-transitioning from a large graph to a significantly smaller one. To address this, we introduce a multidimensional approach utilizing edge expanding.

\subsection{Expanding for $d = 1$}

As shown in Alg. \ref{alg:algorithm6}, we tailor for the scenario where dimensionality $d=1$. A straightforward method is to sort the edges of the initial graph by their significance. The iteration involves progressively incorporating the edge with the maximal value into an empty graph until it evolves into an ($\alpha$,$\beta$)-core encompassing all queries. However, its efficiency is curtailed due to the addition of merely one edge per iteration. To mitigate this, we introduce Lemma~\ref{lemma2} to establish a pseudo-upper bound for $f_1$ within the ESC, allowing batch processing.

\begin{algorithm}[ht]
\caption{EDim1}
\label{alg:algorithm6}
\KwIn{$G,\alpha,\beta,q,d$}
\KwOut{$R$}
\BlankLine
$G^*, C^*, R \gets \varnothing$, $count \gets \alpha \text{ or } \beta$;
$f \gets count^{th}\text{-max } X_1^i \text{ of } q\text{'s edges}$;

\While{$|E(G)| \geq |E(G^*)|$}{
$G^* \gets \{e | X_1^i > f\}$;
$C^* \gets \text{connected component in } G^*$;

\If{$C^* \text{ changed} \land C^* \text{ satisfies Lemmas 3 \& 4}$}{
\If{$q \in C^*$}{
$H \gets (\alpha,\beta)\text{-core of } C^*$;
\If{$q \in H$}{
$R \gets H$;
\textbf{break};
}
}
}
$count \gets count + 1$;
$f \gets count^{th}\text{-max } X_1^i \text{ in } G$;
}
\Return{$R$};
\end{algorithm}

\noindent \textbf{Completeness:}
The existence of an ESC $H'$ with $f_1(H')\geq f_1(H)$, $H'$ must belong to one of the following two cases: (1) it contains edges that are not in $H$, or (2) the edge set of $H'$ is a subset of the edge set of $H$. Since the algorithm continuously adds edges to an empty graph, the resulting community is the first ($\alpha$, $\beta$)-core that includes the query. If $H'$ contains edges that are not in $H$, these edges must have significance less than $f_1(H)$, and if $H'$ contains such edges, it contradicts the definition of $f_1(H')$. So, the first case does not exist. As for (2), if the edge set of $H'$ is a subset of the edge set of $H$, then $H'$ is not the largest ($\alpha$,$\beta$)-core, contradicting the definition of ESC, so there are no other ESCs.

\begin{lem}\label{lemma2}
Given the query vertex $q$, the degree on the layer of $q$ is $\alpha$, the set of significance for edges connected to $q$ store in $N_E$. The pseudo upper bound $\delta$ of the significance of the ESC is the $\alpha^{th}$ maximal value in the sorted set $N_E$.
\end{lem}

By applying Lemma~\ref{lemma2} to determine the pseudo upper bound $\delta$, it used to incorporate only those edges where $X_1^e\geq \delta$, thereby markedly enhancing time efficiency.
If the edges recently integrated fail to constitute an ($\alpha$,$\beta$)-core encompassing the queries, the process necessitates continued iterations, sequentially incorporating remaining edges with the maximum significance into the current graph.
Furthermore, we integrate insights from Lemma~\ref{lemma3} and Lemma~\ref{lemma4} \cite{wang2021efficient}, to further curtail the number of iterations and amplify efficiency. This culminates in the introduction of Algorithm peeling ($d$=1), a novel approach specifically devised to tackle the ESC search problem.

\begin{lem}\label{lemma3}
If a connected subgraph $C^*$ contains an ESC $H$, then there must be: $\alpha\beta - \alpha - \beta \leq|E(C^*)| - |U(C^*)| - |L(C^*)|.$
\end{lem}

\begin{lem}\label{lemma4}
If a connected subgraph $C^*$ contains an ESC $H$, then
$C^*$ has at least $\beta$ vertices of degree $\alpha$ and $\alpha$ vertices of degree $\beta$.
\end{lem}

\noindent \textbf{Time Complexity:} Assuming the graph $G$ has $m$ edges and $n$ vertices, the algorithm initially requires sorting the edges based on significance, with the time complexity of $O(m\cdot \log m)$. Simultaneously, the time complexity to determine whether the current $G$ can generate an ESC is $O(m+n)$, requiring a maximum of $m$ iterations. Consequently, the overall time complexity is $O(m\cdot \log m +m\cdot (m+n))$.

\subsection{Expanding for $d = 2$}

When the edge significance dimension is set to 2, multiple ESCs may exist. Identifying all ESCs requires both edge addition and deletion. We employ a dimensionality reduction approach, processing the two significance dimensions separately. The initial ESC community $H$ is derived using Algorithm expending($d=1$) based on the second-dimensional attribute. By Lemma~\ref{lemma1}, $f_1(H)$ is the lower bound of the first-dimensional significance for all ESCs. A new ESC $H'$ with $f_1(H') > f_1(H)$ is then found by removing edges with $X_1^e = f_1(H)$ and adding those with $X_1^e > f_1(H)$. This iterative process continues until all ESCs are identified, forming the basis of Algorithm expending ($d=2$). As shown in Alg. \ref{alg:algorithm7}, it initializes key variables: $G^*$ (accumulated edges), $f_1$ (current ESC's first-dimensional significance), $R$ (set of ESCs), $count$ (tracking edge order by second-dimensional significance), and $D$ (operating dimension). ESC identification consists of two steps: finding the first ESC and generating subsequent ones.

\begin{algorithm}[ht]
\caption{EDim2}
\label{alg:algorithm7}
\KwIn{$G,\alpha,\beta,q,F$}
\KwOut{$R$}
\BlankLine
$G^*, R \gets \varnothing$, $f_1 \gets -1$, $count \gets \alpha \text{ or } \beta$, $D \gets 1$;
$f \gets count^{th}\text{-max } X_2^i \text{ of } q\text{'s edges}$;

\While{$|E(G^*)| < |E(G)|$}{
\eIf{$f_1 = -1$}{
$G^* \gets \{e | X_2^i > f\}$;
$D \gets 1$, $count \gets count + 1$;
$f \gets count^{th}\text{-max } X_2^i \text{ in } G$;
}{
$G^* \gets (G^* \setminus \{e | X_1^i = f_1\}) \cup \{e | X_1^i > f_1\}$;
$D \gets 2$;
}

$S \gets \varnothing$;
\While{$H \gets (\alpha,\beta)\text{-core of } G^* \neq \emptyset \text{ and } q \in H \text{ and } F \subseteq H$}{
$S \gets H$;
Remove edge with $\min X_D^i$ from $G^*$;
}

\uIf{$S = \varnothing$}{\lIf{$f_1 = -1$}{continue}\textbf{break}}
\uElseIf{$\exists \text{ESC} \in R \preceq S$}{Remove dominated ESC from $R$}

$(f_1,f_2) \gets S$;
$R \gets R \cup \{(f_1,f_2)\}$;
}
\Return{$R$};
\end{algorithm}

\begin{figure}[ht]
\centering
\begin{subfigure}{0.32\columnwidth}
\includegraphics[width=\linewidth]{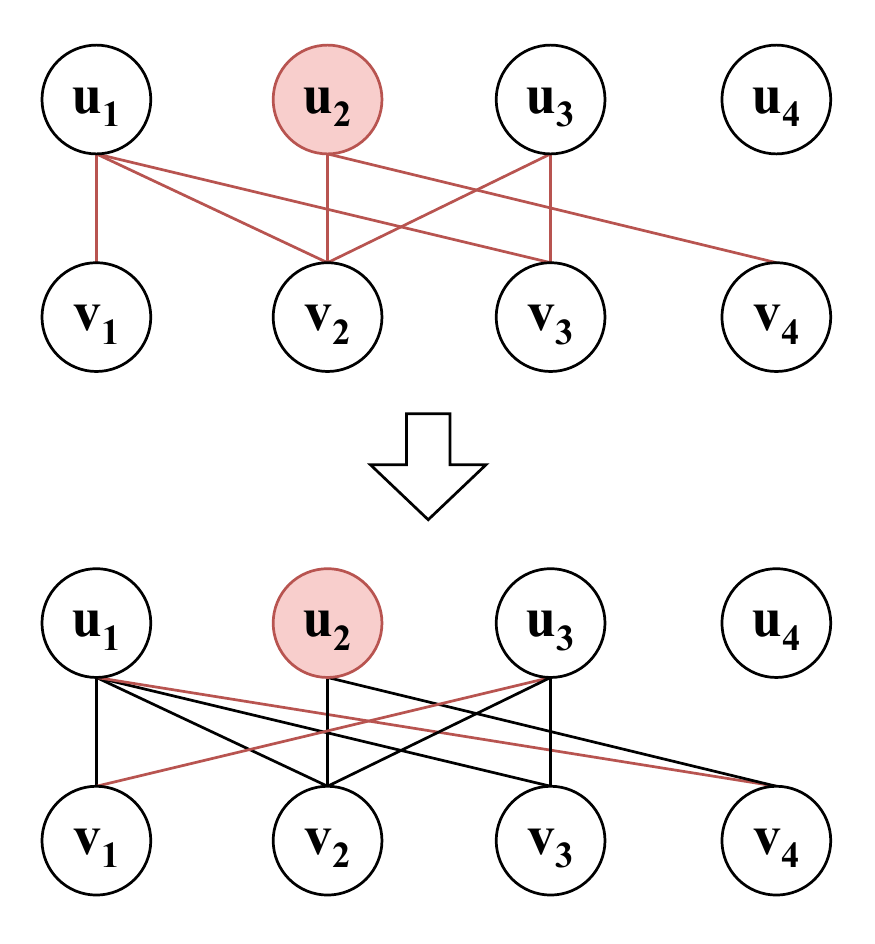}
\caption{}
\label{fig:expandingd=2:case_a}
\end{subfigure}
\begin{subfigure}{0.32\columnwidth}
\includegraphics[width=\linewidth]{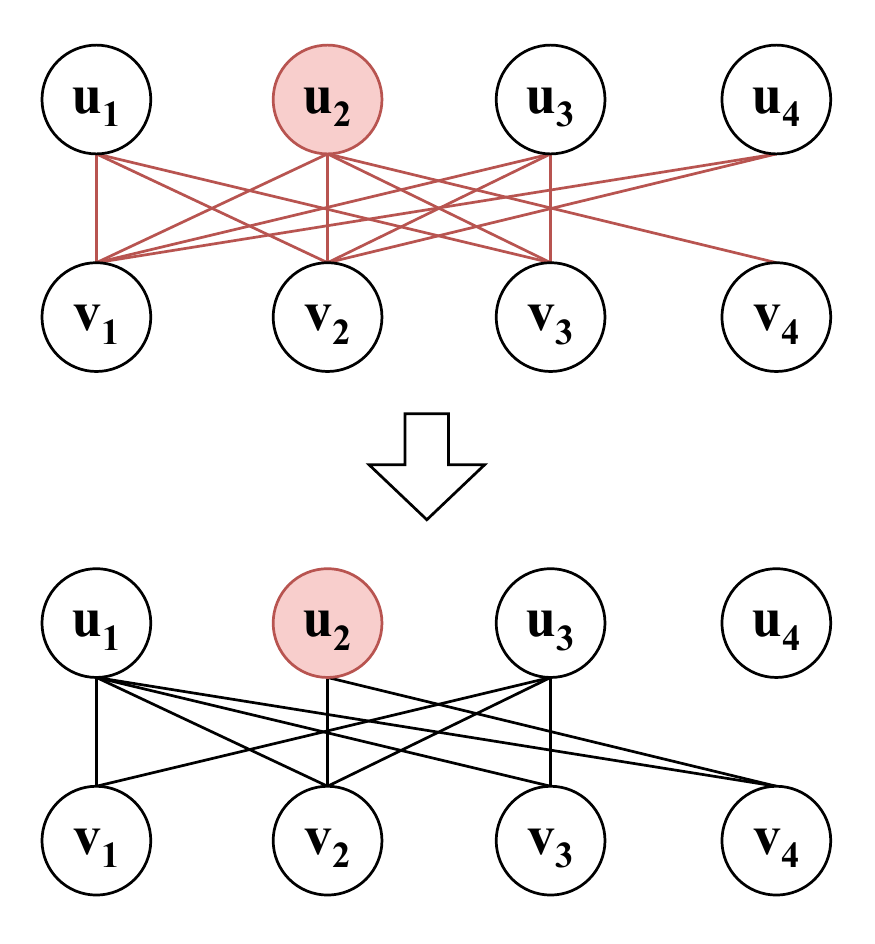}
\caption{}
\label{fig:expandingd=2:case_b}
\end{subfigure}
\begin{subfigure}{0.32\columnwidth}
\includegraphics[width=\linewidth]{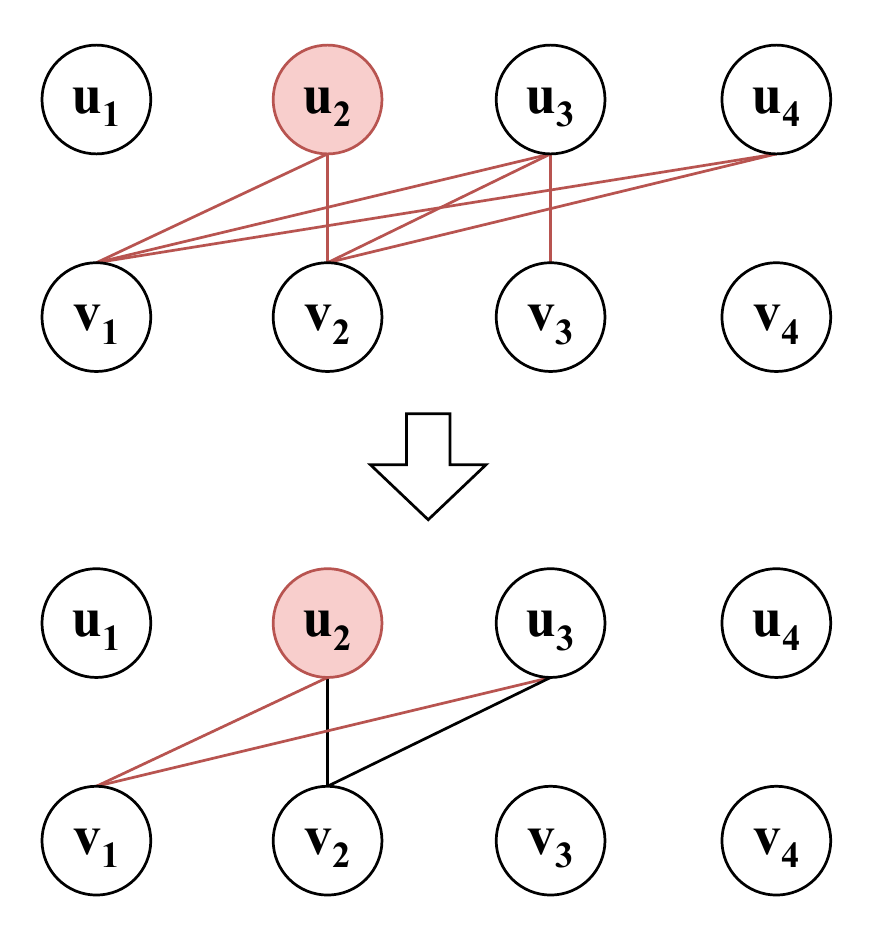}
\caption{}
\label{fig:expandingd=2:case_c}
\end{subfigure}
\caption{An example of expanding in case $d=2$.}
\label{fig:expandingd=2}
\end{figure}

\begin{example}
Using the example in Fig~\ref{fig1} with query vertex $U_2$ and $\alpha = \beta = 2$, we apply Algorithm peeling($d$=2) to find all ESCs, as shown in Fig~\ref{fig:expandingd=2}. First, let $f$ be the second-largest second-dimensional significance of edges connected to $U_2$, initially 10. Edges with $X_2^e > 10$ are added to $G^*$ (upper part of Fig~\ref{fig:expandingd=2:case_a}). Updating $f$ to 9, we continue adding edges until $f=7$, forming the first ESC (1,7). The lower part of Fig~\ref{fig:expandingd=2:case_a} highlights the additional edges in red. Next, removing edge ($u_1$, $v_4$) and adding edges with $X_1^e > 1$ to $G^*$, we iteratively delete the edge with the smallest second-dimensional significance to obtain ESC (2,6) (Fig~\ref{fig:expandingd=2:case_b}). Repeating this process generates ESCs (3,6), (4,6), and (8,6), but due to dominance relationships, the final ESCs are (1,7) and (8,6) (Fig~\ref{fig:expandingd=2:case_c}).
\end{example}

\noindent \textbf{Time Complexity:} Next, we analyze the time complexity of the Algorithm~\ref{alg:algorithm7}. The algorithm initially requires sorting the edges based on significance, with a time complexity of $O(m\cdot \log m)$. Next, the time complexity for obtaining the ($\alpha$,$\beta$)-core is $O(m+n)$, and during the removing edges phase, approximately $m$ times ($\alpha$,$\beta$)-core operations are required. Considering the need for $m$ edge addition operations, the overall time complexity of the algorithm is $O(m\cdot \log m+m^2 \cdot(m+n))$.

\subsection{Expanding for $d = 3$}

Expanding strategies can still be extended to three-dimensional situations. Utilizing a dimensionality reduction strategy, the values of the three-dimensional attributes are efficiently decomposed into two distinct sets: one set of one-dimensional values and another set of two-dimensional values. A preliminary and simplistic strategy entails enumerating all feasible values of \( f_3 \), setting each of these values consecutively, and subsequently conducting a subgraph analysis to locate ESCs within the context of the remaining two-dimensional attributes.

\begin{figure}[ht]
\centering
\begin{subfigure}{0.3\columnwidth}
\includegraphics[width=\linewidth]{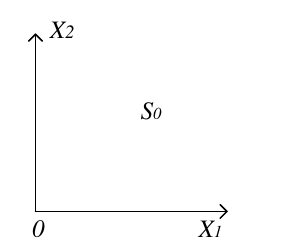}
\caption{}
\label{fig:recursivedecomposition:case2_1}
\end{subfigure}
\begin{subfigure}{0.3\columnwidth}
\includegraphics[width=\linewidth]{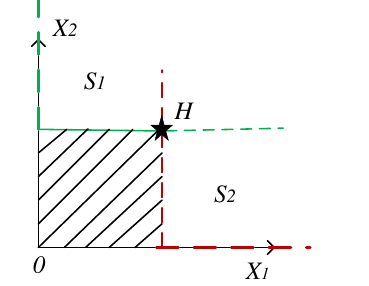}
\caption{}
\label{fig:recursivedecomposition:case2_2}
\end{subfigure}
\caption{Recursive decomposition approach.}
\label{fig:recursivedecomposition}
\end{figure}

However, this method fails to guarantee finding ESCs for every \( f_3 \) value. Fixing \( f_3 \) constrains the search domain to \( C_0 = \{X_1^e > 0, X_2^e > 0\} \), as shown in Fig.~\ref{fig:recursivedecomposition:case2_1}. An ESC \( H \) defines a rectangular region in the solution space (shaded area in Fig.~\ref{fig:recursivedecomposition:case2_2}). A new ESC within this region would be subordinate to \( H \), leading to redundant computations and reduced efficiency.

To optimize the search, we use recursive decomposition, pruning unnecessary computations. An ESC must lie outside the shaded area, but its irregular shape prevents direct application of Algorithm peeling($d$=2). Thus, we partition the solution space into \( C_1 = \{X_1^e > 0, X_2^e > f_2(H)\} \) and \( C_2 = \{X_1^e > f_1(H), X_2^e > 0\} \), enabling separate applications of Algorithm peeling($d$=2). Each region's lower bounds are represented as \((0, f_2(H))\) for \( C_1 \) and \((f_1(H), 0)\) for \( C_2 \).

\begin{lem}\label{lemma5}
Suppose it exists $n$ ESCs: $ H_i = (f_1(H_i),f_2(H_i))$, $i \in [1, n]$,
arranging all the horizontal and vertical coordinates as:
\begin{equation}
\begin{gathered}
0<f_1(H_1) < f_1(H_2) < ...<f_1(H_n)\\
	 f_2(H_1) > f_2(H_2)>...> f_2(H_n)>0
\end{gathered}
\end{equation}
By combining the coordinates of the upper and lower inequalities, the solution space is:
$C_1=(0,f_2(H_1))$,$C_2=(f_1(H_1),f_2(H_2))$,...,$C_{n+1}=(f_1(H_n),0)$.
\end{lem}

As shown in Alg. \ref{alg:algorithm8}, we combine the aforementioned recursive decomposition manner for solving the ESC search problem in three-dimensional attribute bipartite graphs, which initially creates an empty graph $G^*$ for each solution space. Subsequently, it iteratively adds a batch of edges to $G^*$ that meet the significance requirements of the solution space. The algorithm then calls Algorithm peeling($d$=2) on $G^*$ to compute ESCs. Following this, leveraging Lemma~\ref{lemma5}, it updates the solution space using the generated ESCs and calculates $f_3$ for each solution space. This process iterates until all ESCs are identified.

\begin{algorithm}[ht]
\caption{EDim3}
\label{alg:algorithm8}
\KwIn{$G,\alpha,\beta,q,F$}
\KwOut{$R$}
\BlankLine
$G^*, R, Q, P \gets \varnothing$;
$Q \gets Q \cup (0,0,\text{EDim1}(G,\alpha,\beta,q,3))$;

\While{$Q \neq \emptyset$}{
$f_3 \gets \max_{3\text{-dim}} Q$;
$S \gets \varnothing$;

\ForEach{$(p_1,p_2,f_3) \in Q$}{
$G^* \gets \{e | x_1^e > p_1, x_2^e > p_2, x_3^e \geq f_3\}$;
Remove $(p_1,p_2,f_3)$ from $Q$;
\lIf{$x_3^e = f_3$}{$F \gets F \cup e$}

$S \gets S \cup \text{EDim2}(G^*,\alpha,\beta,q,F)$;

\ForEach{$(f_1,f_2) \in S$}{
\uIf{$(f_1,f_2) \succeq P$}{Remove dominated from $P$;}
\uElseIf{$(f_1,f_2) \preceq P$}{continue;}
$Q \gets Q \cup (f_1,f_2,f_3)$;
$P \gets P \cup (f_1,f_2)$;
}

\ForEach{$(point_1,point_2) \in \text{Lemma5}$}{
$G^* \gets \{e | x_1^e > point_1, x_2^e > point_2\}$;
$Q \gets Q \cup (point_1,point_2,\text{EDim1}(G^*,\alpha,\beta,q,3))$;
}
}
}
\Return{$R$};
\end{algorithm}

\noindent \textbf{Time Complexity:} Assuming there are a total of $\epsilon$ ESCs, the algorithm will invoke Algorithm~\ref{alg:algorithm7} $\epsilon$ times and Algorithm~\ref{alg:algorithm6} $\epsilon$ times. At the same time, we also need $\epsilon^2$ times the solution space division. Consequently, the overall time complexity is $O(\epsilon \cdot m^2 \cdot (m+n)+\epsilon^2+\epsilon\cdot m\cdot(m+n))$.

\subsection{Universal Expanding}

Dimensionality reduction can be used for high-dimensional attributes. As shown in Alg. \ref{alg:algorithm9}, we propose Algorithm peeling ($d>3$) for the ESC search problem, following the methodology of Algorithm peeling ($d=3$) but extending the solution space update in recursive decomposition based on Lemma~\ref{lemma5}.

\begin{algorithm}[ht]
\caption{UniEDim}
\label{alg:algorithm9}
\KwIn{$G,\alpha,\beta,q,F,d$}
\KwOut{$R$}
\BlankLine
\lIf{$d=3$}{\Return{\text{EDim3}(G,$\alpha$,$\beta$,q,F)}}
$G^* \gets \emptyset$; $R \gets \emptyset$; $Q \gets \emptyset$; $P \gets \emptyset$;
$f_d^{max} \gets \text{EDim1}(G,\alpha,\beta,q,d)$;
$Q \gets Q \cup \{(0,\dots,0, f_d^{max})\}$;
$C \gets \{(0,\dots,0)_{d-1}\}$;

\While{$Q \neq \emptyset$}{
$f_d \gets \max_{d\text{-dim}} Q$;
$S \gets \emptyset$;
\ForEach{$(p_1,\dots,p_{d-1},f_d) \in Q$}{
$G^* \gets \{e \mid x_1^e > p_1, \dots, x_{d-1}^e > p_{d-1}, x_d^e \geq f_d\}$;
Remove $(p_1,\dots,p_{d-1},f_d)$ from $Q$;
\lIf{$x_d^e = f_d$}{$F \gets F \cup e$}

$S' \gets \text{UniEDim}(G^*,\alpha,\beta,q,F,d-1)$;
$S \gets S \cup S'$;

\ForEach{$(f_1,\dots,f_{d-1}) \in S$}{
\uIf{$(f_1,\dots,f_{d-1}) \succeq P$}{Remove dominated from $P$;}
\uElseIf{$(f_1,\dots,f_{d-1}) \preceq P$}{\textbf{continue};}

$Q \gets Q \cup \{(f_1,\dots,f_d)\}$;
$P \gets P \cup \{(f_1,\dots,f_{d-1})\}$;
$C \gets \text{DivideSpace}(C, (f_1,\dots,f_{d-1}))$;
}

\ForEach{$(point_1, \dots, point_{d-1}) \in C$}{
$G^* \gets \{e \mid x_1^e > point_1, \dots, x_{d-1}^e > point_{d-1}\}$;
$f_d^{new} \gets \text{EDim1}(G^*,\alpha,\beta,q,d)$;
$Q \gets Q \cup \{(point_1,\dots,point_{d-1}, f_d^{new})\}$;
}
}
}
\Return{$R$};

\SetKwFunction{FMain}{DivideSpace}
\SetKwProg{Fn}{Function}{:}{end}
\Fn{\FMain{$C, (f_1,\dots,f_k)$}}{
\ForEach{$i \in \{1, \dots, k\}$}{
$C' \gets \emptyset$;
\ForEach{$c \in C$}{
\uIf{$(f_1,\dots,f_k) \preceq c$}{
$c' \gets (c_1,\dots,c_{i-1},f_i,\dots,c_k)$;
Remove $c$ from $C$;
$C' \gets C' \cup \{c'\}$;
}
}
$C \gets C \cup C'$;
}
\Return{$C$};
}
\end{algorithm}

\noindent \textbf{Time Complexity:} Assume that there are $\epsilon$ number of ESC solutions. Algorithm~\ref{alg:algorithm9} will be iterated approximately $d-1$ times. In each iteration process, it needs to be called $\epsilon$ times, so the overall time complexity is $O(\epsilon^{d-2}\cdot m^2\cdot(m+n)+\epsilon^{d-1}+\epsilon^{d-2}\cdot m(m+n))$.

\section{Experiment}
We validate the efficiency by varying the dimension $d$ and structural parameters $\alpha$ and $\beta$. To assess scalability, we change the dataset size. Additionally, we demonstrate the algorithm's effectiveness through a case study. All experiments are conducted on an Ubuntu server with a 2.40GHz Intel(R) Xeon(R) Gold 6240R CPU and 512GB memory in Python. Table~\ref{tab:datasets_parameters} provides the statistics of datasets, of which $d_{1m}$, $d_{2m}$, $d_{1a}$ and $d_{2a}$ denote the maximal degree of the upper vertices, the maximal degree of the lower vertices, the average degree of the upper vertices and the average degree of the lower vertices, respectively. Table \ref{tab:time} provides the comprehensive time complexity analysis.

\begin{table}[ht]
\centering
\caption{Datasets and Parameters used in our experiments}
\begin{minipage}{0.68\columnwidth}
\centering
\resizebox{\textwidth}{!}{
\begin{tabular}{|c|c|c|c|c|c|c|}
\hline
Dataset & Vertices & Edges & $d_{1m}$ & $d_{2m}$ & $d_{1a}$ & $d_{2a}$\\
\hline
Crime & 1K & 1K & 25 & 18 & 1.8 & 2.7\\
\hline
arXiv & 39K & 59K & 116 & 18 & 3.5 & 2.6\\
\hline
DBpedia & 225K & 294K & 28 & 12K & 1.7 & 5.5\\
\hline
BookCrossing & 446K& 1.1M & 14K & 2.5K & 10.9 & 3.4\\
\hline
IMDB & 872K & 2.7M & 654 & 1.3K & 3.9 & 14.6\\
\hline
TV Tropes & 152K & 3.2M & 6.5K & 12K & 50.2 & 36.9\\
\hline
MovieLens & 222K & 25.0M & 32.2K & 81.4K & 154.3 & 423.7 \\
\hline
\end{tabular}}
\end{minipage}
\hfill
\begin{minipage}{0.28\columnwidth}
\centering
\resizebox{\textwidth}{!}{
\begin{tabular}{|c|c|}
\hline
Para. & Tested values \\ \hline
$d$ & 1, 2, 3, 4 \\ \hline
$\alpha(\beta)$ & 1, 2, 3, 4 \\ \hline
$\sigma$ & 20, 40, 60, 80, 100\% \\ \hline
\end{tabular}}
\end{minipage}
\label{tab:datasets_parameters}
\end{table}

\begin{table}[ht]
\centering
\caption{Time Complexity of ESC Algorithm}
\label{tab:time}
\resizebox{\columnwidth}{!}{
\begin{tabular}{c|c|c}
\toprule
Dim & Peeling & Expanding\\
\midrule
$d=1$ & $O(m (n+m))$ & $O(m\cdot \log m +m(m+n))$ \\
$d=2$ & $O(\epsilon \cdot(m-m/n)(n+m))$ & $O(m\cdot \log m+m^2(m+n))$ \\
$d=3$ & $O(m\cdot(m-m/n)(n + m))$ & $O(\epsilon \cdot m^2 (m+n)+\epsilon^2+\epsilon\cdot m(m+n))$ \\
$d>3$ & $O(m^{d-1}\cdot(m-m/n)(n + m))$ & $O(\epsilon^{d-2}\cdot m^2(m+n)+\epsilon^{d-1}+\epsilon^{d-2}\cdot m(m+n))$\\
\bottomrule
\end{tabular}}
\end{table}

\textbf{Datasets.}
We collected seven datasets, i.e., DBpedia, BookCrossing, IMDB, TV Tropes, arXiv, Crime, and MovieLens from the KONECT website. These datasets are all undirected bipartite graphs, and neither edges nor vertices have associated significance. As the currently available datasets lack edge weight attributes, we generated non-negative numerical multidimensional attributes for all attribute-less bipartite graph datasets, ensuring these attributes were independent. We used a uniform distribution for this purpose because it generates values with equal probability and avoids the correlation biases inherent in the Gaussian distribution. In real-world scenarios, edges with high significance in one dimension rarely exhibit high significance in others, making \textit{anti-correlation} more common and justifying the use of uniform distribution.

\textbf{Parameters.} We vary three parameters: dimension ($d$), structural parameter ($\alpha$), and dataset completeness ($\sigma$). Since $\alpha$ and $\beta$ similarly constrain the community structure, we vary only $\alpha$. Table~\ref{tab:datasets_parameters} lists the parameter ranges and default values. For each experiment, we randomly selected 20 query vertex sets from each graph that ensured the existence of the maximal ($\alpha$, $\beta$)-core, and averaged the time required for these sets.

\subsection{Efficiency evaluation}
\begin{figure*}[ht]
\centering
\begin{minipage}{0.13\textwidth}
\centering
\includegraphics[width=\linewidth]{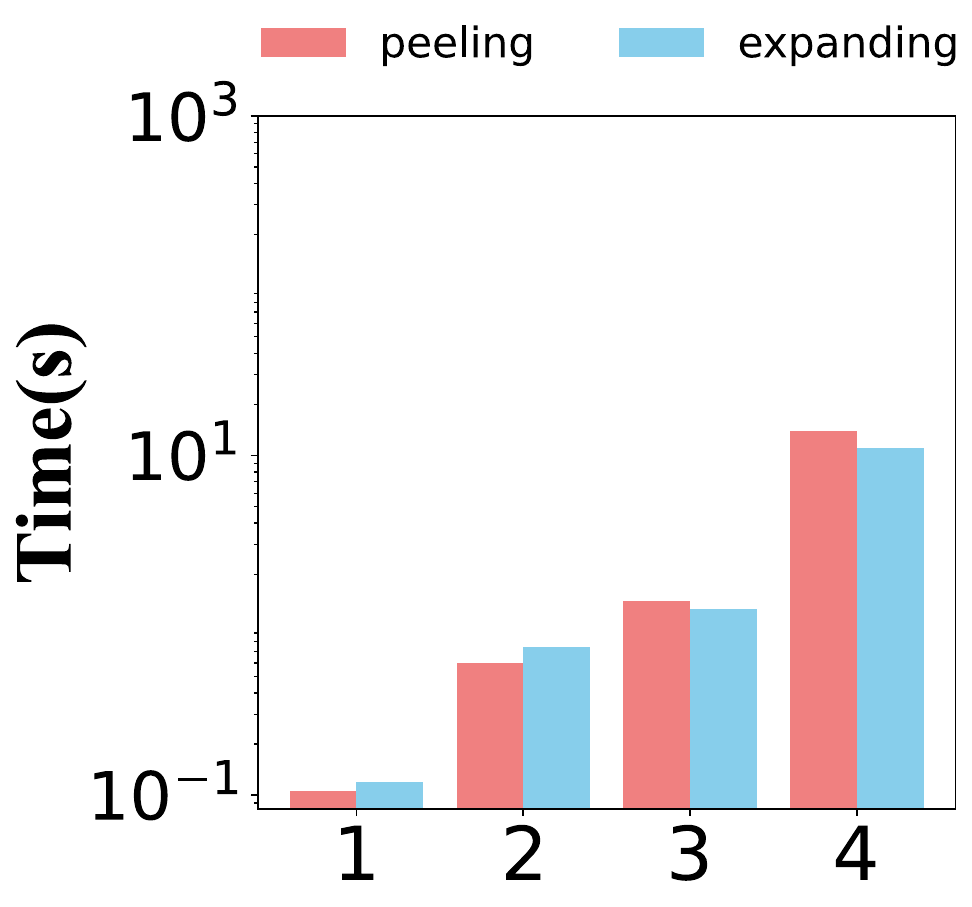}
\subcaption{Crime}
\label{fig:Crime-dim}
\end{minipage}
\hfill
\begin{minipage}{0.13\textwidth}
\centering
\includegraphics[width=\linewidth]{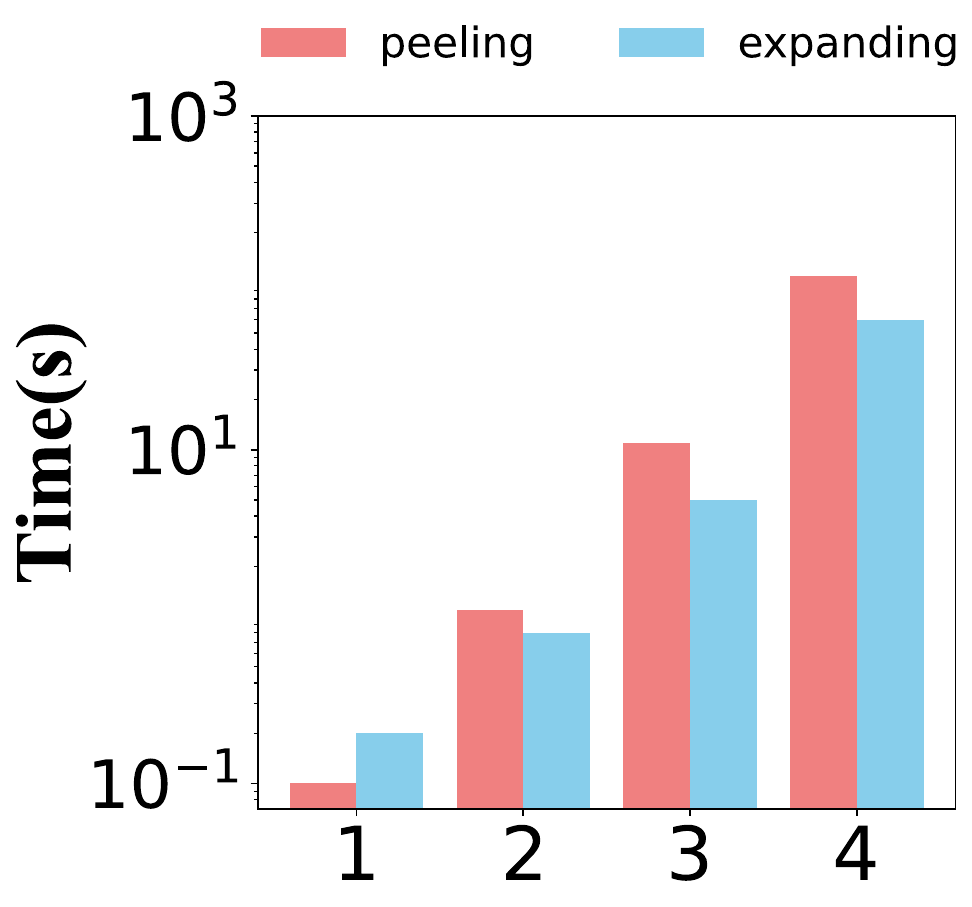}
\subcaption{arXiv}
\label{fig:arXiv-dim}
\end{minipage}
\hfill
\begin{minipage}{0.13\textwidth}
\centering
\includegraphics[width=\linewidth]{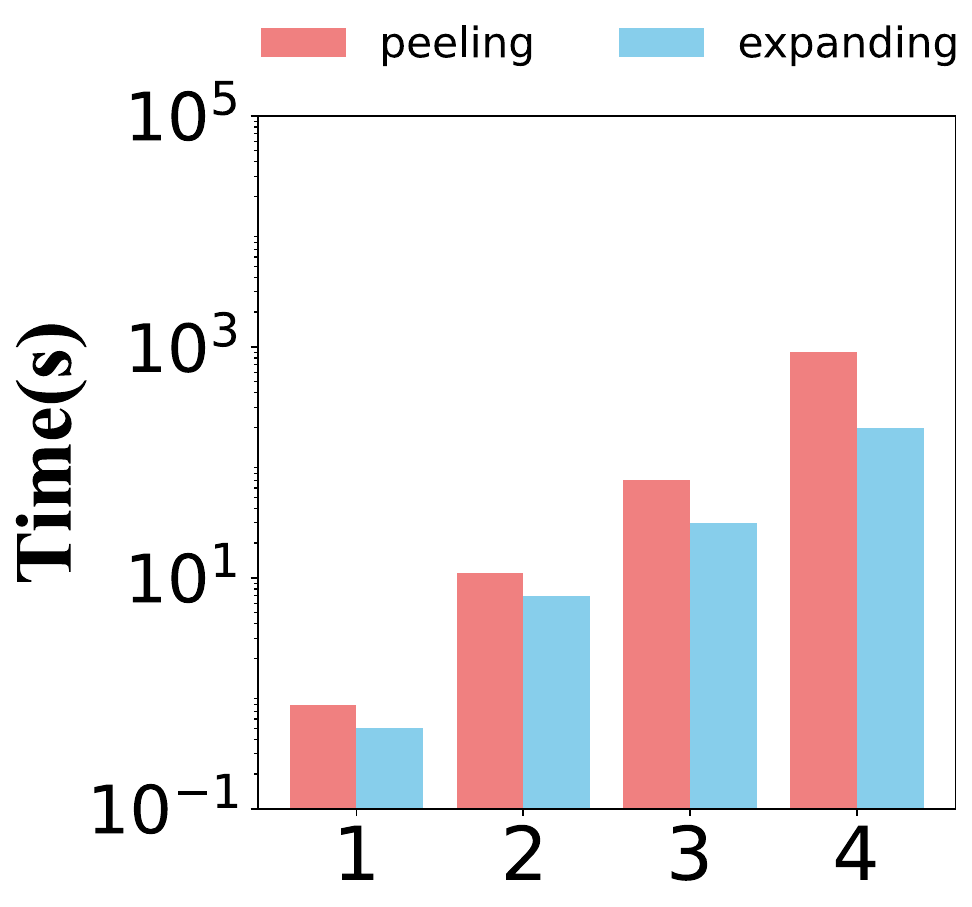}
\subcaption{DB}
\label{fig:dbpedia-dim}
\end{minipage}
\hfill
\begin{minipage}{0.13\textwidth}
\centering
\includegraphics[width=\linewidth]{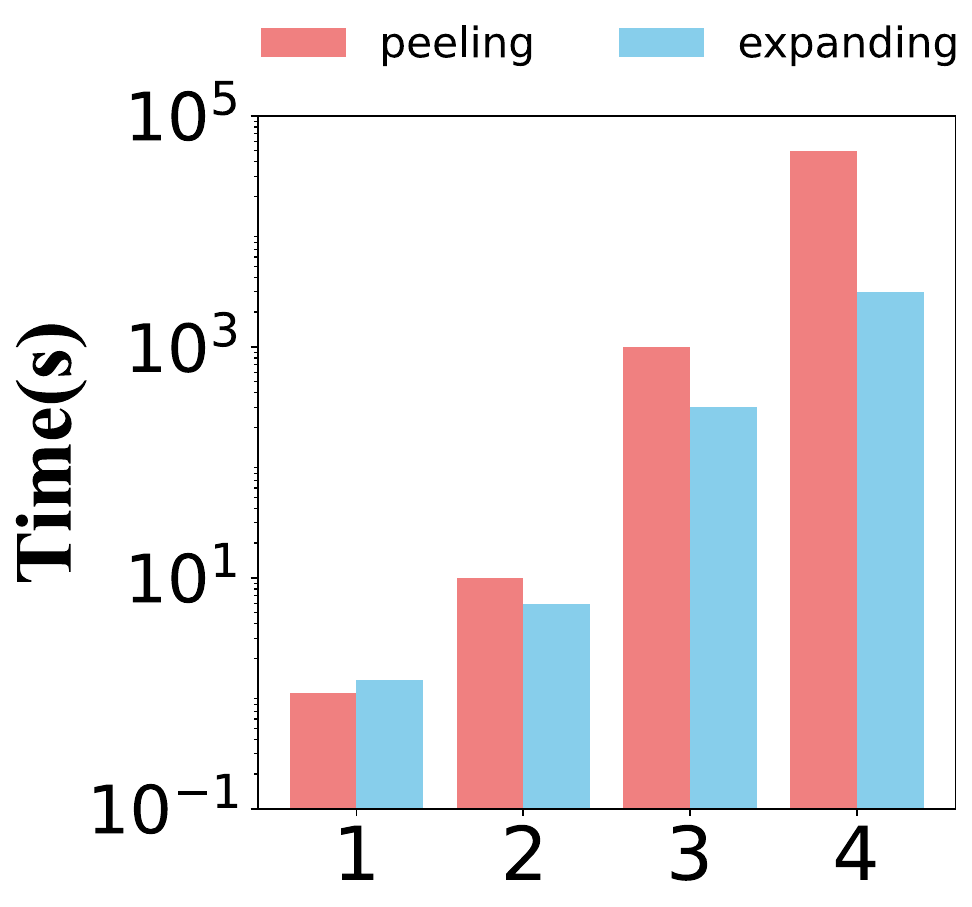}
\subcaption{BC}
\label{fig:bookcrossing-dim}
\end{minipage}
\hfill
\begin{minipage}{0.13\textwidth}
\centering
\includegraphics[width=\linewidth]{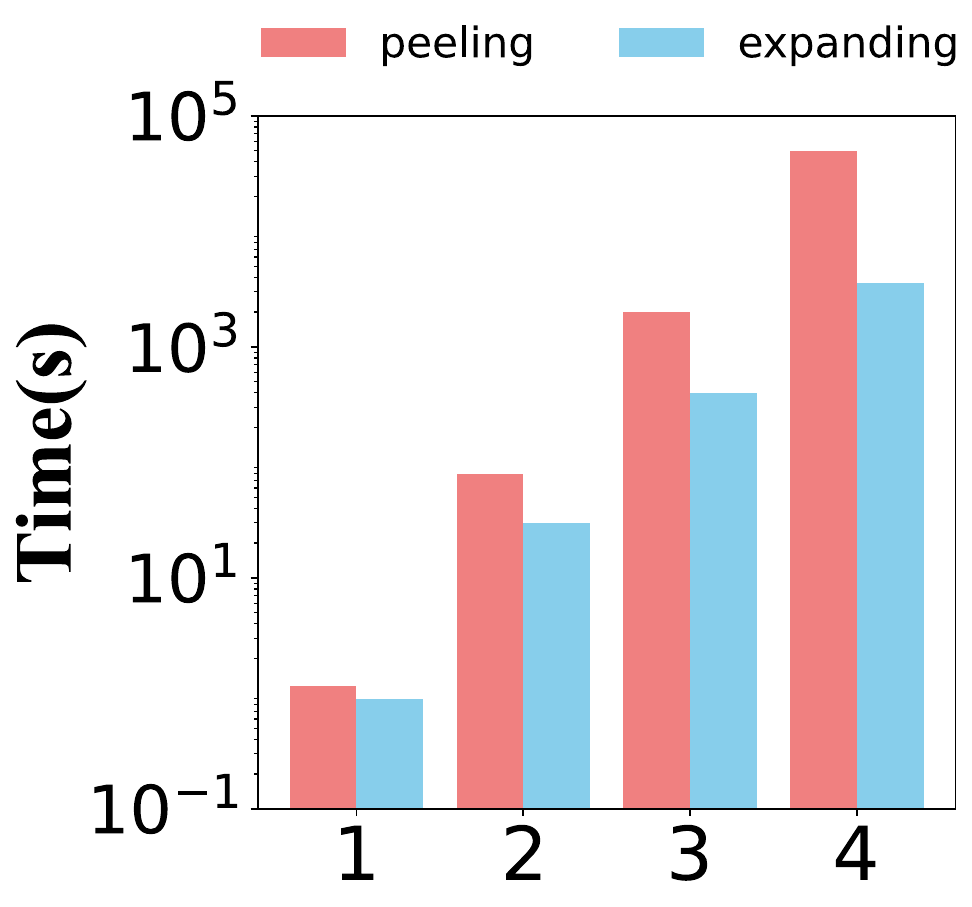}
\subcaption{IMDB}
\label{fig:imdb-dim}
\end{minipage}
\hfill
\begin{minipage}{0.13\textwidth}
\centering
\includegraphics[width=\linewidth]{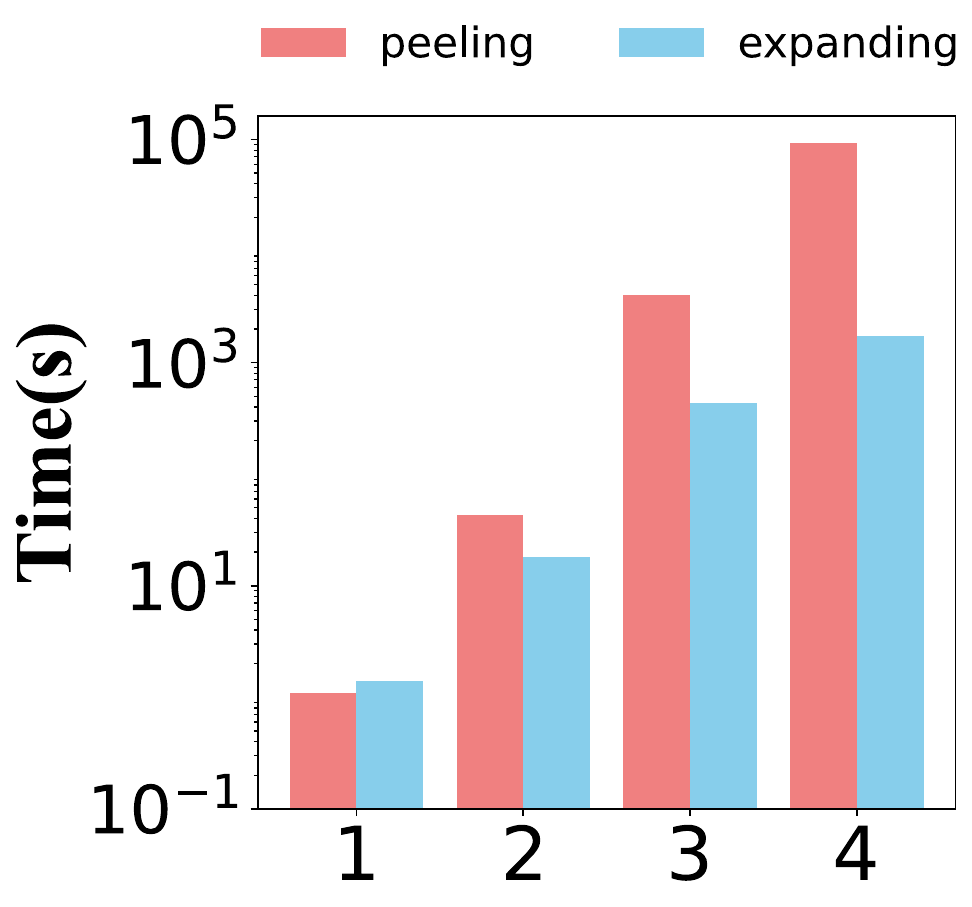}
\subcaption{TVT}
\label{fig:tvtropes-dim}
\end{minipage}
\hfill
\begin{minipage}{0.13\textwidth}
\centering
\includegraphics[width=\linewidth]{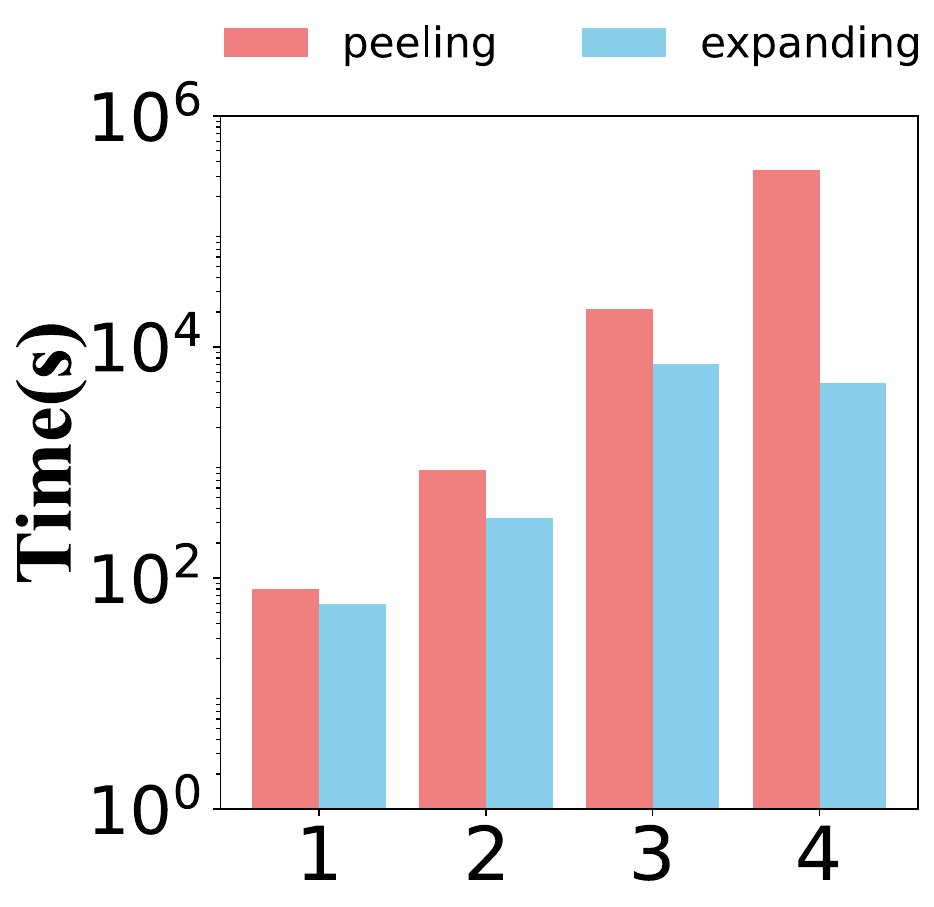}
\subcaption{ML}
\label{fig:MovieLen-dim}
\end{minipage}
\caption{The efficiency of the algorithms w.r.t dimensions change.}
\label{fig6}
\end{figure*}

We investigate the time required to search for ESCs across various datasets by varying the attribute dimensionality $d$ while keeping other variables constant. As shown in Fig~\ref{fig6}, we fix the structural constraint parameters $\alpha$ and $\beta$ at 2 and use the average time taken for 20 query vertices as the search time for each dimensionality. Note that, we use DB to denote DBpedia, BC to denote BookCross, TVT to denote TVTropes, ML to denote MovieLen. Fig~\ref{fig6}(a)-(g) indicates that as attribute dimensionality increases, the time required for the peeling and expanding algorithms also increases.

\subsection{Effectiveness evaluation}

\textbf{Scalability evaluation.} Without loss of generality, we take the expanding algorithm as the experimental object. To assess the scalability of Algorithm peeling ($d=3$), we randomly select 20\%, 40\%, 60\%, 80\%, and 100\% of the edges from the input dataset (vary $\sigma$) to form subgraphs for our experiments. The experimental results confirm that the algorithm consistently produces correct outcomes. As shown in Fig~\ref{fig7}, the algorithm's runtime generally rises with an increasing number of edges. However, deviations occur due to query vertex selection. For example, if a 20\% edge subgraph fully covers a query vertex's ESC, adding more edges doesn't raise the processing time for that vertex.

\begin{figure*}[ht]
\centering
\begin{minipage}{.13\textwidth}
\centering
\includegraphics[width=\linewidth]{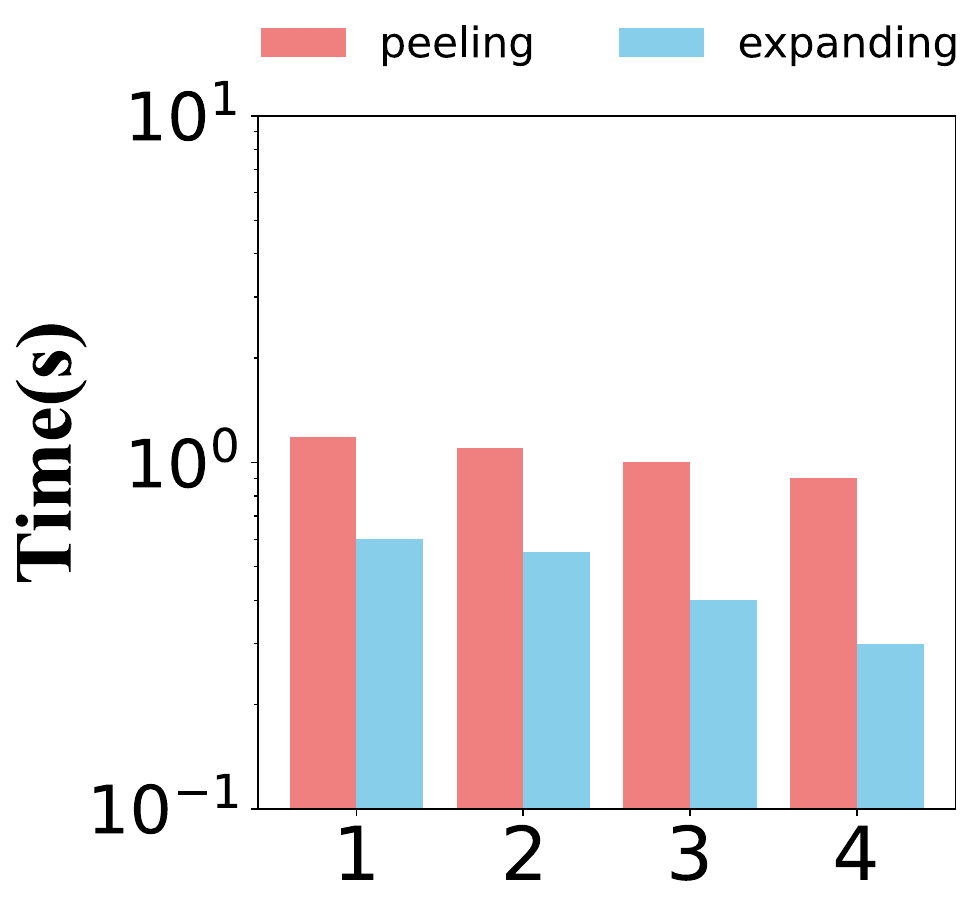}
\subcaption{Crime}
\label{fig:Crime-alpha}
\end{minipage}
\hfill
\begin{minipage}{.13\textwidth}
\centering
\includegraphics[width=\linewidth]{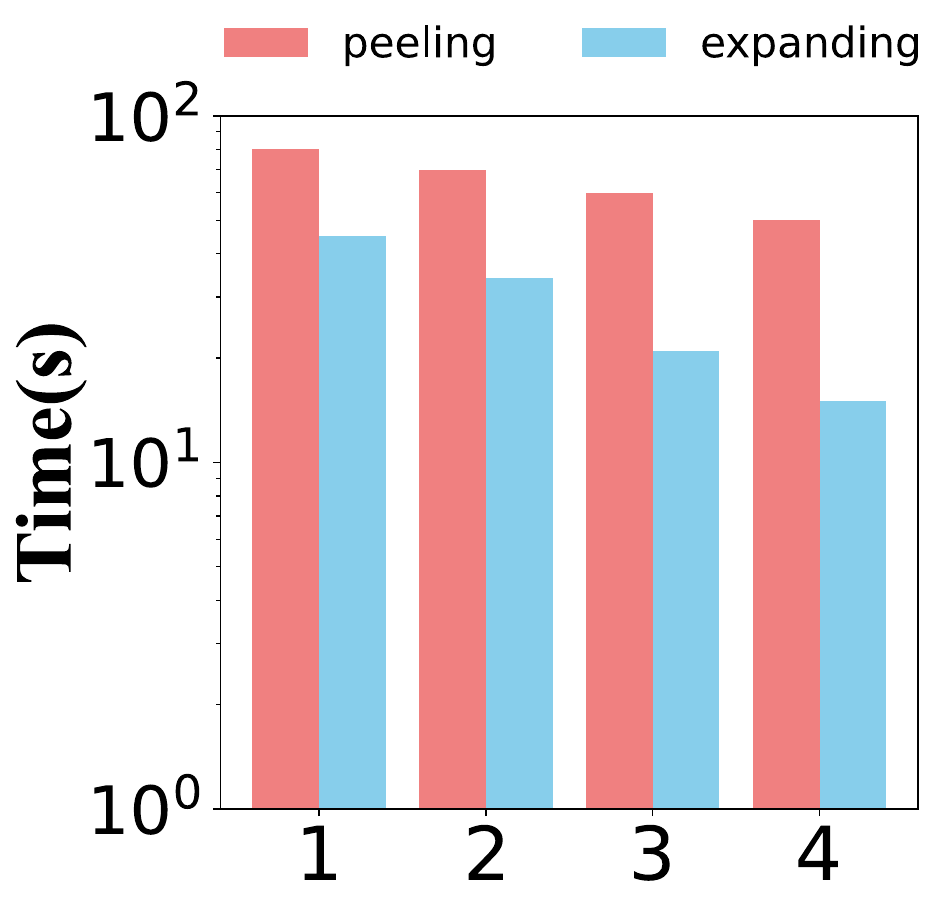}
\subcaption{arXiv}
\label{fig:arXiv-alpha}
\end{minipage}
\hfill
\begin{minipage}{.13\textwidth}
\centering
\includegraphics[width=\linewidth]{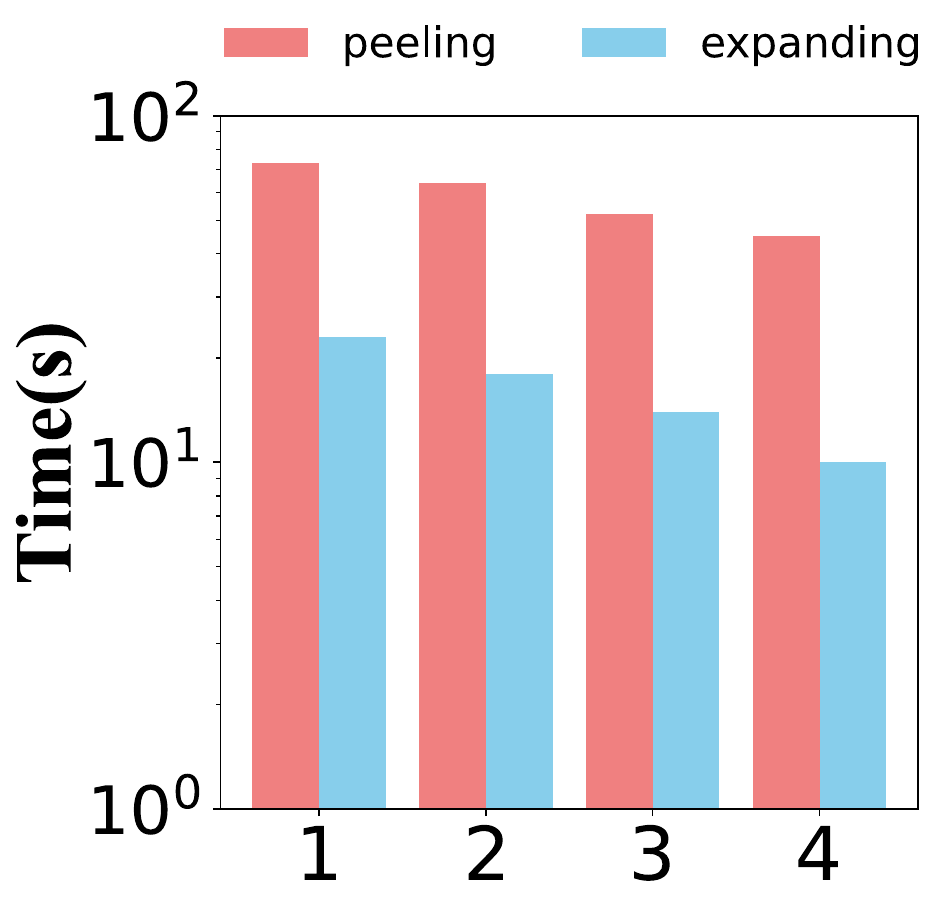}
\subcaption{DB}
\label{fig:dbpedia-alpha}
\end{minipage}
\hfill
\begin{minipage}{.13\textwidth}
\centering
\includegraphics[width=\linewidth]{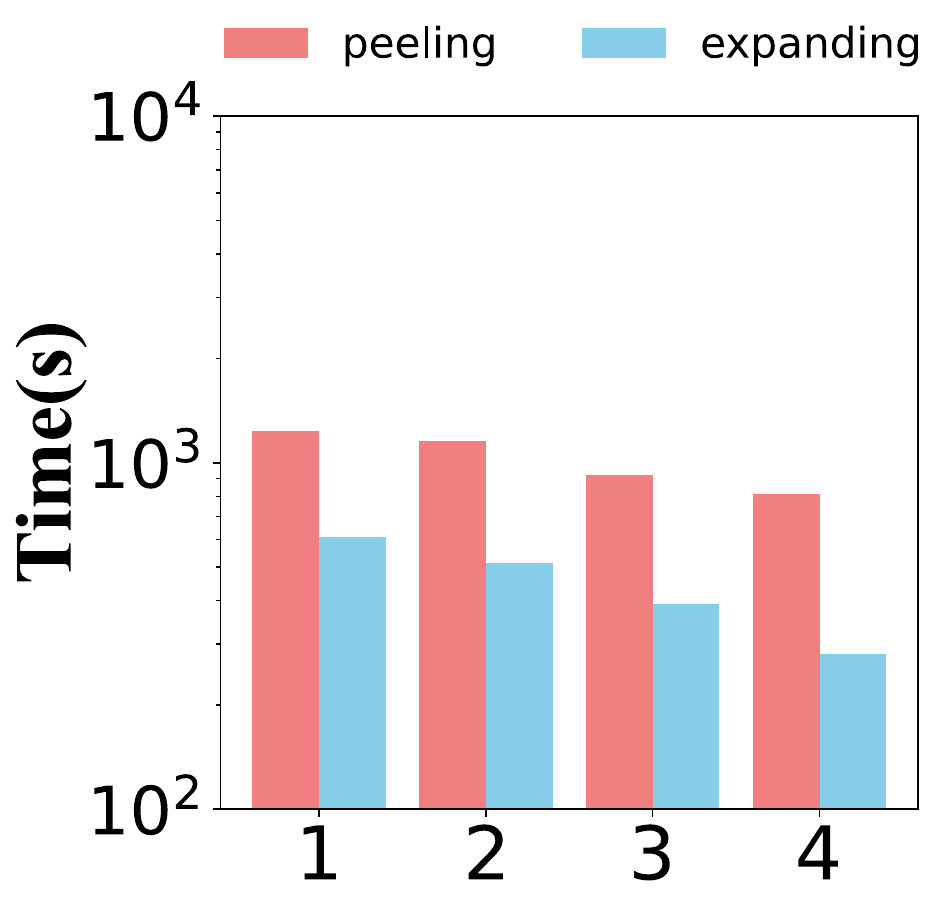}
\subcaption{BC}
\label{fig:bookcrossing-alpha}
\end{minipage}
\hfill
\begin{minipage}{.13\textwidth}
\centering
\includegraphics[width=\linewidth]{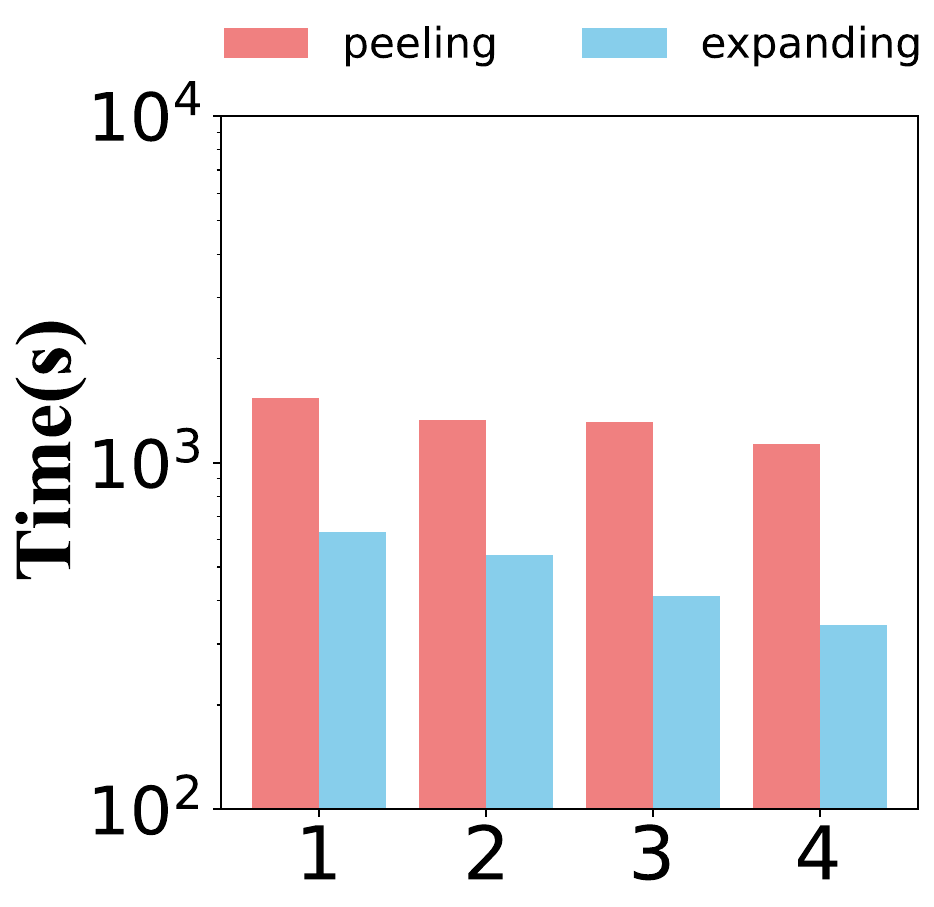}
\subcaption{IMDB}
\label{fig:imdb-alpha}
\end{minipage}
\hfill
\begin{minipage}{.13\textwidth}
\centering
\includegraphics[width=\linewidth]{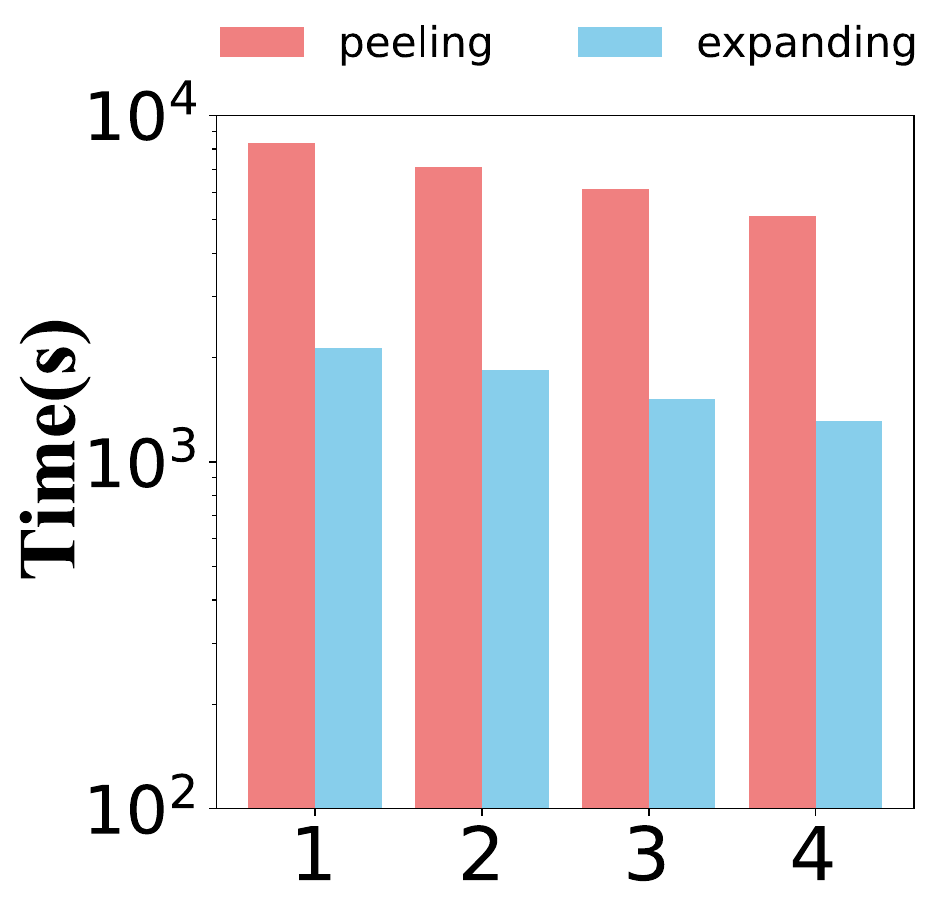}
\subcaption{TVT}
\label{fig:tvtropes-alpha}
\end{minipage}
\hfill
\begin{minipage}{.13\textwidth}
\centering
\includegraphics[width=\linewidth]{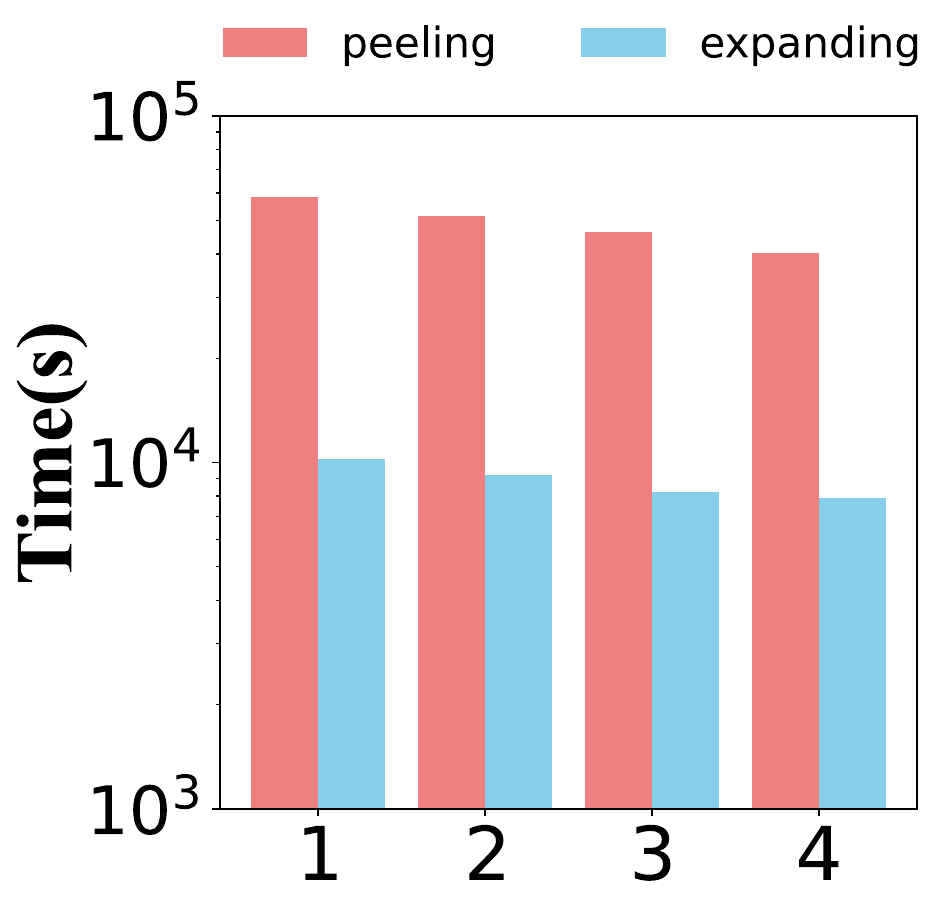}
\subcaption{ML}
\label{fig:MovieLen-alpha}
\end{minipage}
\caption{The effectiveness of the algorithms w.r.t $\alpha$ changes.}
\label{fig8}
\end{figure*}

\begin{figure*}[ht]
\centering
\begin{minipage}{.4\textwidth}
\centering
\includegraphics[width=\linewidth]{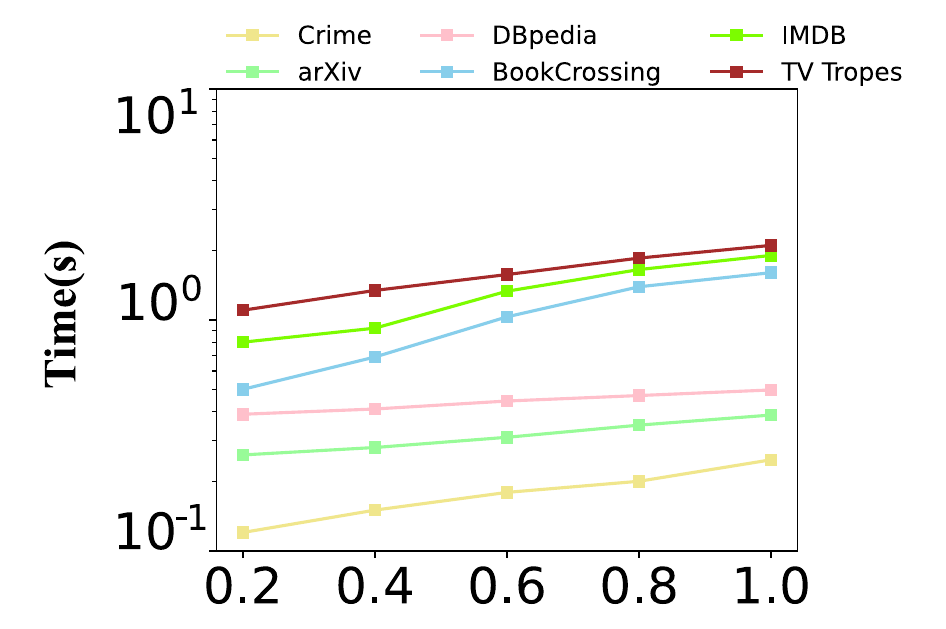}
\subcaption{$d=1$}
\label{fig:d1_scab}
\end{minipage}
\hfill
\begin{minipage}{.4\textwidth}
\centering
\includegraphics[width=\linewidth]{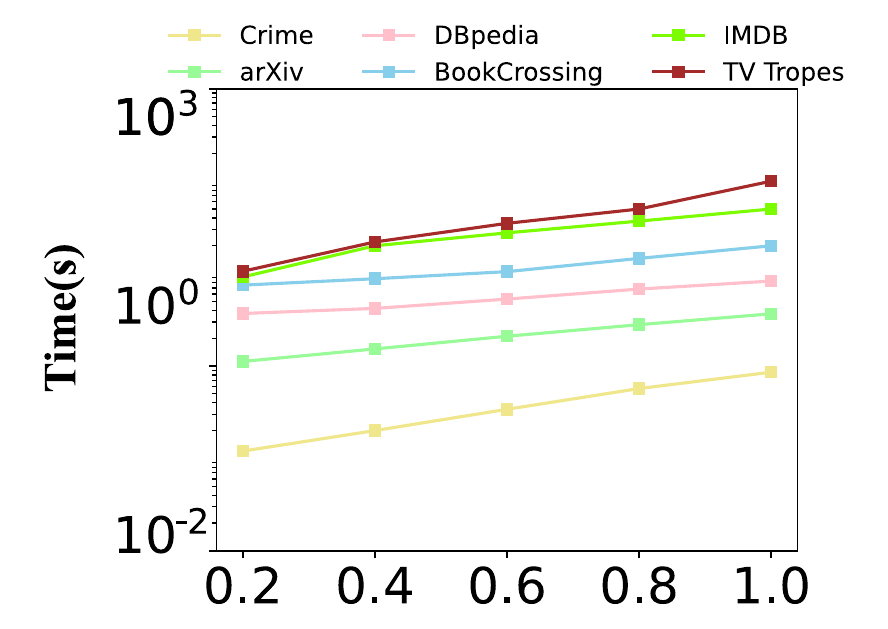}
\subcaption{$d=2$}
\label{fig:d2_scabv}
\end{minipage}

\begin{minipage}{.4\textwidth}
\centering
\includegraphics[width=\linewidth]{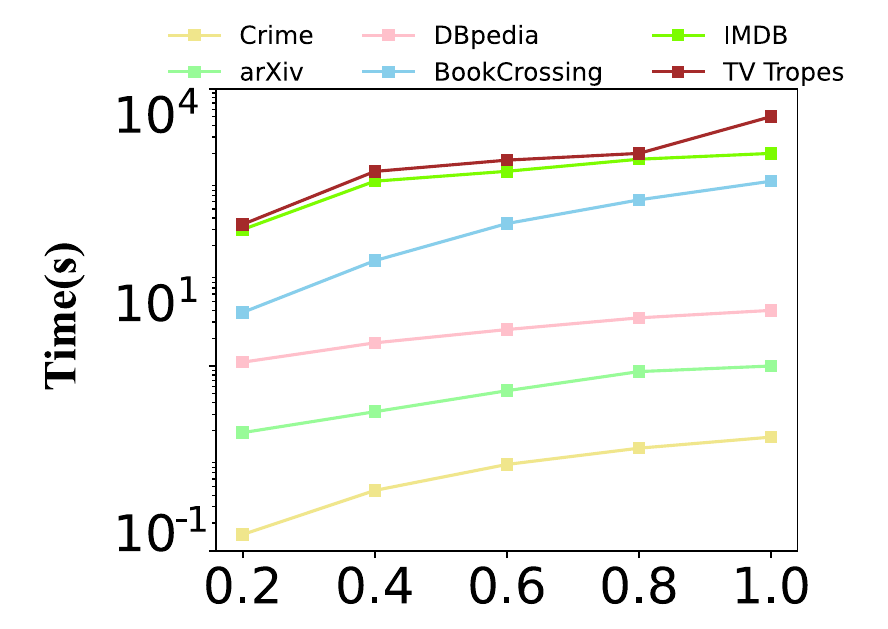}
\subcaption{$d=3$}
\label{fig:d3_scab}
\end{minipage}
\hfill
\begin{minipage}{.4\textwidth}
\centering
\includegraphics[width=\linewidth]{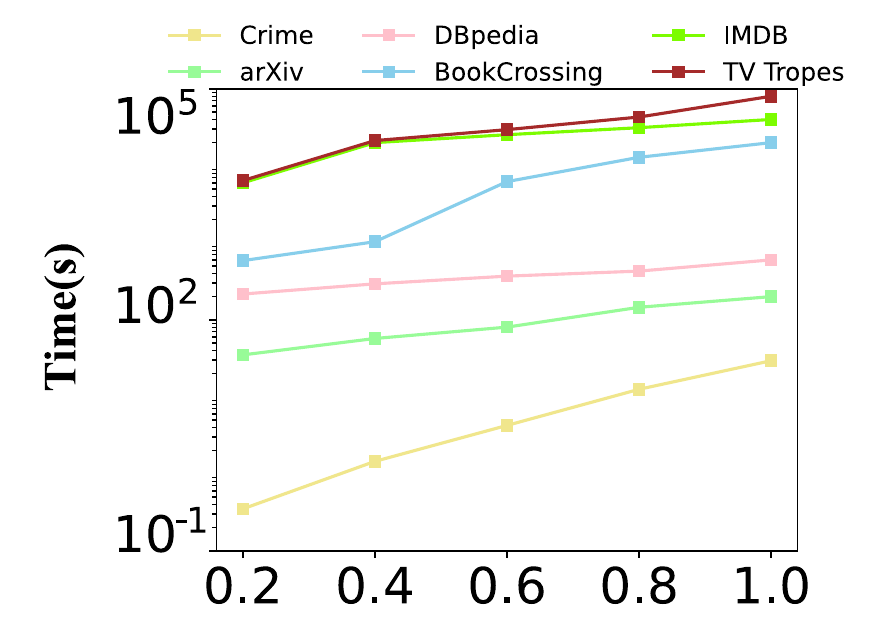}
\subcaption{$d=4$}
\label{fig:d4_scab}
\end{minipage}

\caption{Scalability test w.r.t varing dimensions.}
\label{fig7}
\end{figure*}

\noindent \textbf{Evaluating the effect of $\alpha$.} We maintain other variables constant and vary the value of $\alpha$ across different datasets. As depicted in Fig~\ref{fig8}(a)-(g), both the peeling and expanding algorithms exhibit a decreasing trend in search time as $\alpha$ increases. Note that, we use DB to denote DBpedia, BC to denote BookCross, TVT to denote TVTropes, ML to denote MovieLen. This trend is due to the fact that as $\alpha$ increases, the size of the initial maximal ($\alpha$,$\beta$)-core decreases, which in turn reduces the time required by the algorithms. Additionally, we observe that the efficiency of the expanding algorithm gradually surpasses that of the peeling algorithm. This is because a higher $\alpha$ leads to a larger discrepancy between the size of the ($\alpha$,$\beta$)-core and the original graph, resulting in the expanding algorithm requiring less time.

\subsection{Case study}

\begin{figure}[ht]
\centering
\includegraphics[width=1\columnwidth]{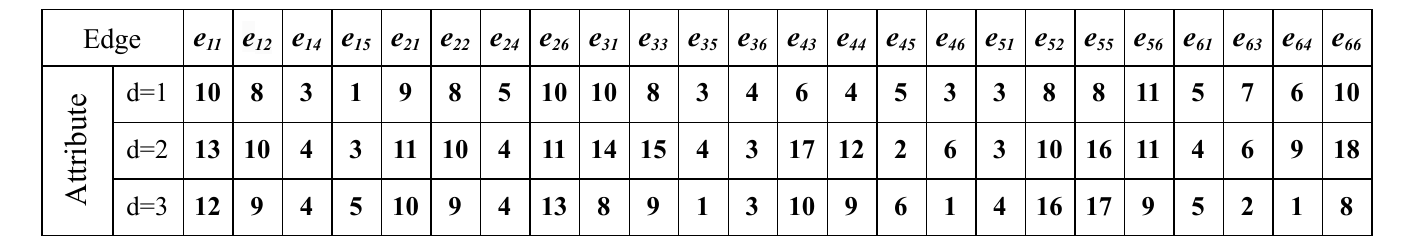}
\begin{subfigure}{0.48\columnwidth}
\includegraphics[width=\linewidth]{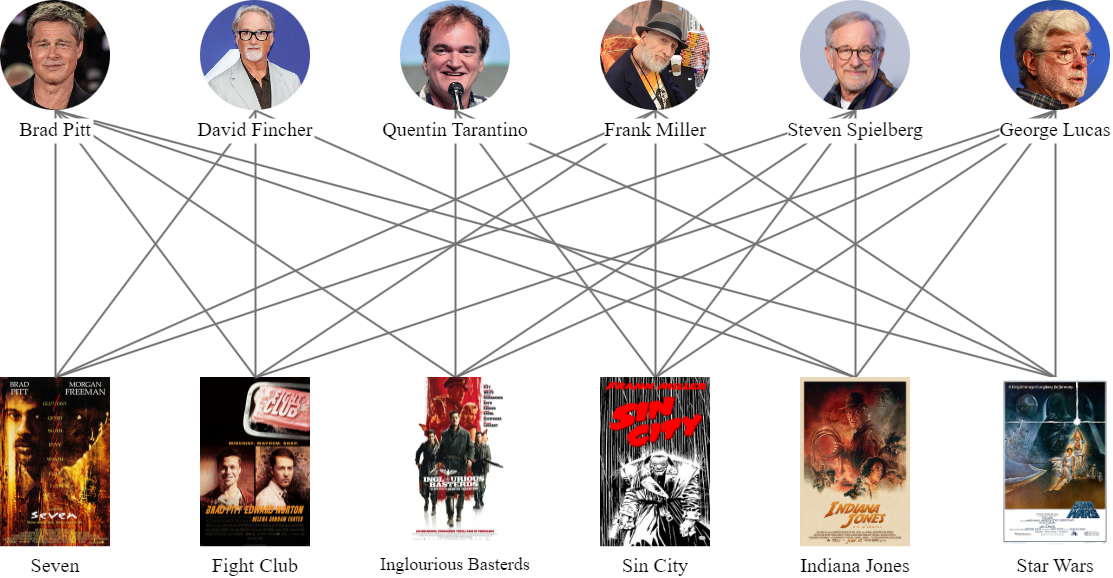}
\caption{\scriptsize{initial state.}}
\end{subfigure}
\hfill
\begin{subfigure}{0.48\columnwidth}
\includegraphics[width=\linewidth]{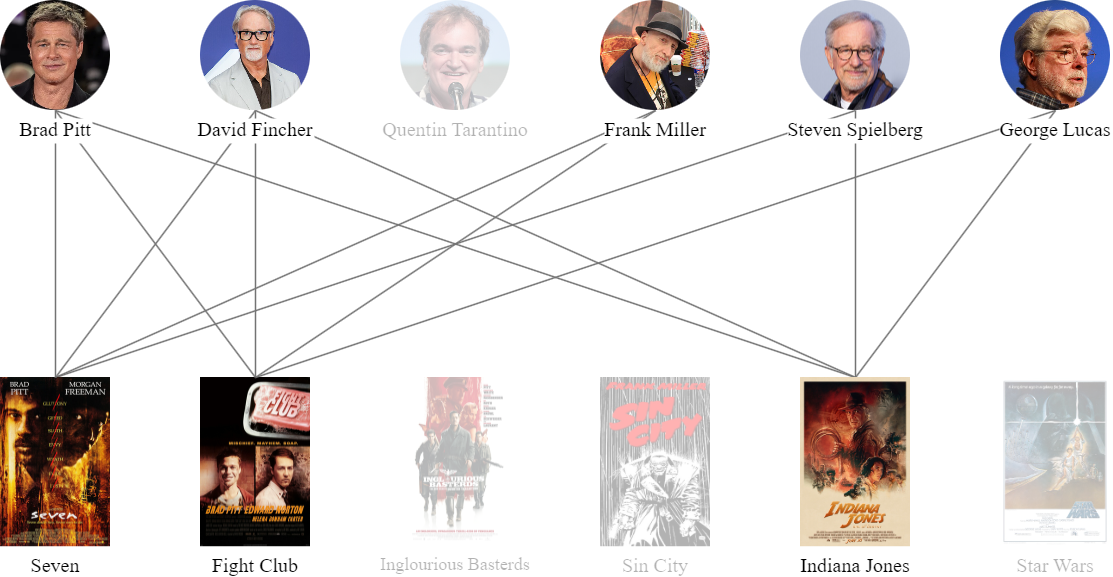}
\caption{\scriptsize{($\alpha$, $\beta$)-core.}}
\label{fig:case_b}
\end{subfigure}
\begin{subfigure}{0.48\columnwidth}
\includegraphics[width=\linewidth]{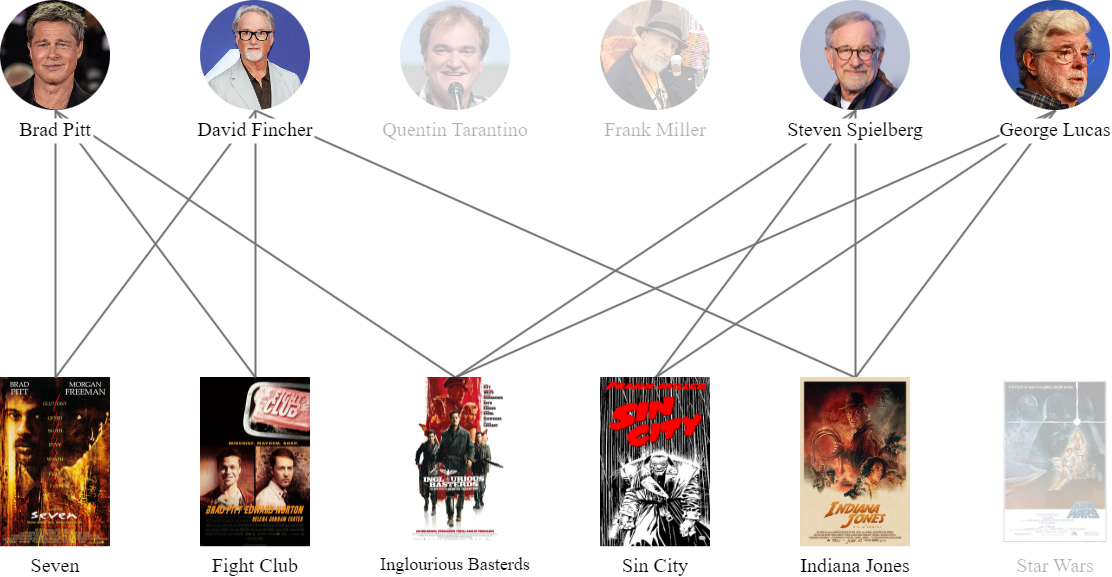}
\caption{\tiny{Significant($\alpha$, $\beta$).}}
\label{fig:case_c}
\end{subfigure}
\hfill
\begin{subfigure}{0.48\columnwidth}
\includegraphics[width=\linewidth]{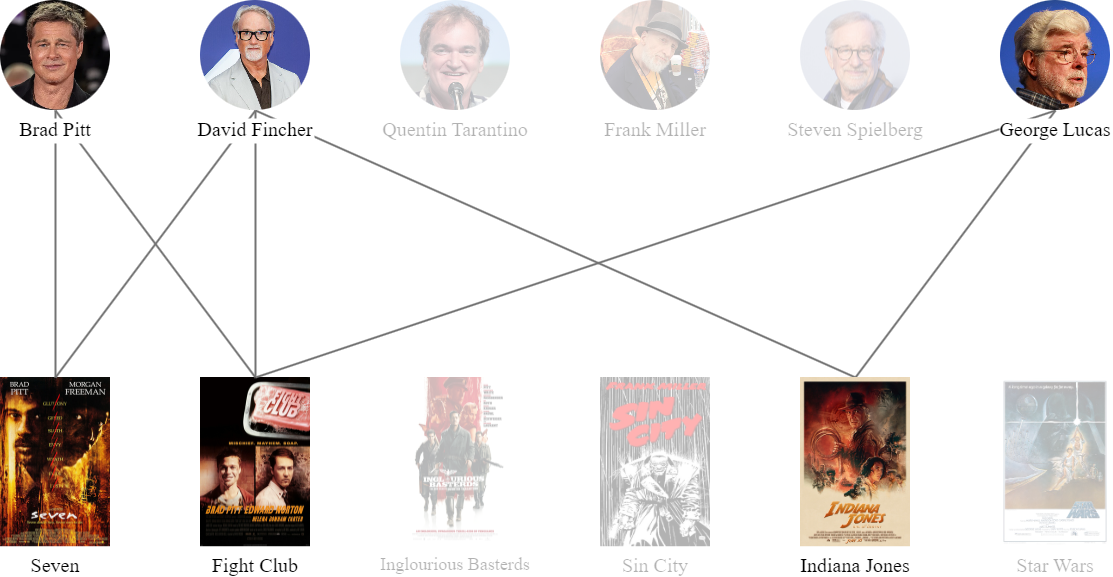}
\caption{\scriptsize{ESC.}}
\label{fig:case_d}
\end{subfigure}

\caption{Case study: comparison with different models.}
\label{fig4}
\end{figure}

\textbf{Comparison with different models.}
As shown in Figure~\ref{fig4}, we carried out a case study on a subset of the IMDB dataset. Given that each edge in the dataset lacks attribute information, ``$e_{ij}$'' represents the edge between the $i$-th movie and the $j$-th director, we assigned three-dimensional attribute values to each edge. The first dimension signifies the duration of a director's performance, the second dimension denotes the director's remuneration, and the third dimension represents the director's rating. All three dimensions encompass synthetic attribute information. Figure~\ref{fig4}(a) showcases the initial state of the community. When we focus on the performance of a single dimension (where $d$=$1$) and query vertex for \emph{Brad Pitt} with parameters $\alpha = 2$ and $\beta = 4$,the communities identified by the algorithm proposed by \cite{liu2023distributed} are illustrated in Figure~\ref{fig4}(b). In contrast, the communities identified by the algorithm proposed by \cite{wang2021efficient} for the first dimension alone are shown in Figure~\ref{fig4}(c). The ESCs found by our algorithm are displayed in Figure~\ref{fig4}(d). The case study's outcomes clearly indicate that our algorithm's ESCs have a denser structure compared to those identified by \cite{liu2023distributed}. Our algorithm effectively filters out edges with minimal attribute values in the second dimension, such as $e_{35}$ and $e_{36}$, which highlights its precision in identifying more relevant solutions. Moreover, while \cite{wang2021efficient} identifies an ESC with $f(H_1)$, our algorithm not only recognizes $H_1$ but also uncovers an additional ESC with $f(H_2) = (8,10,9)$. They are non-dominated, each excels in one dimension, underscoring that our design can discover more precise solutions in bipartite graphs with multi-dimensional attribute values than prior art\cite{wang2021efficient}.

\section{Related Work}
\textbf{Community search (CS).}
The objective of CS is to identify cohesive subgraphs related to queries~\cite{cui2014local}, which can be divided into precise algorithms and heuristic algorithms. Precise aim to find communities that meet precise optimization criteria, often involving complex mathematical modeling and calculations. Typically, models such as~\cite{bonchi2019distance} and~\cite{zhang2019unboundedness} are employed to discover precise internally cohesive communities. Heuristic use approximate methods to quickly find better communities. Some representative CS approaches are based on local information. For instance, in approach ~\cite{cui2014local}, the minimum degree of the $k$-core is utilized as a constraint for community discovery \cite{bi2017optimal,chen2016efficient}. Additionally, ~\cite{li2015influential} takes vertex attributes into account by utilizing importance values in the CS process.

\noindent \textbf{Community search on attributed graphs.}
Attributed graphs, categorized into $k$-core~\cite{li2018skyline,chen2018exploring}, $k$-truss~\cite{huang2017attribute,2017Finding}, and $k$-clique~\cite{2017Most} structures, are crucial for CS. \cite{fang2016effective} addressed CS on keyword attribute graphs by defining attribute communities based on structural cohesion and keyword aggregation, introducing the CL-tree index and algorithms. \cite{fang2017effective} developed a maintenance algorithm for the CL-tree index. For location graphs, \cite{fang2017effectivee} introduced spatial-aware communities and provided an exact solution. \cite{fang2018spatial} proposed three fast algorithms for continuous spatial-aware CS on dynamic spatial graphs. In influence attribute graphs, \cite{li2015influential} proposed an online search algorithm for the top-$j$ influential communities with a linear space index. \cite{chen2016efficient,bi2017optimal,li2017finding} introduced algorithms for large graphs and small $j$ values. \cite{guo2021multi} proposed multi-attributed joint community search in road social networks, while \cite{liu2023significant} introduced the HSC model based on meta-path cohesiveness in heterogeneous graphs.

\noindent \textbf{Community search on bipartite graphs.}
Applications of CS in bipartite graphs are largely confined to one-dimensional attributes. \cite{wang2021efficient} addressed the CS problem in edge-weighted bipartite graphs by introducing the significant $(\alpha,\beta)$-community concept, which captures vertex cohesion and maximizes the minimum edge weight within a subgraph. They proposed an indexing structure with $O(\delta \cdot m)$ for community retrieval, along with algorithms for expanding and peeling from queries to identify significant $(\alpha,\beta)$-communities. \cite{2021Pareto} explored the CS problem in influence bipartite graphs, introducing the Pareto-optimal $(\alpha,\beta)$-community model, which considers both subgraph cohesion and vertex importance. It yields Pareto-optimal $(\alpha,\beta)$-communities with vertex degree constraints and integrates Pareto optimality with $O(p \cdot m)$.

\section{Conclusion}
In this paper, we address the challenge of finding edge-attributed skyline communities (ESCs) in large bipartite graphs. We introduce a new model and propose an efficient peeling approach by iteratively removing edges with the least significant attributes across dimensions. A more effective expanding approach is designed to reduce search space, featuring a novel upper bound. Experimental results on real-world large bipartite graphs confirm the effectiveness and efficiency.

\subsubsection{Acknowledgements}
This work was supported by the NSFC (No. 62302485), and the Key Research Project of Chinese Academy of Sciences (No. RCJJ-145-24-21).

\bibliographystyle{splncs04}
\bibliography{sample-base}
\end{document}